\documentclass[12pt, letter]{article}
\usepackage{natbib}
\usepackage{latexsym}
\usepackage{graphics}
\usepackage{amsfonts}
\usepackage{enumerate}
\usepackage{graphicx}
\usepackage{epsf,boxedminipage,lscape}
\usepackage{epstopdf}
\usepackage{epsfig}
\usepackage{verbatim}
\usepackage{subfig}
\usepackage{caption}
\usepackage{footnote}
\usepackage{slashbox}
\usepackage{lscape}
\usepackage{hhline}
\usepackage{array}
\usepackage{graphicx}
\usepackage{setspace}
\usepackage{amssymb}
\usepackage{amsmath}
\usepackage{amsfonts}
\usepackage{times}
\usepackage{color}
\usepackage{indentfirst}
\usepackage{epsfig}
\usepackage{lscape,ctable,rotating}
\usepackage{enumitem}
\setlist{nolistsep}
\usepackage[left=1in,top=1in,right=1in,foot=1in]{geometry}

\newcommand{\R}{\mathbb{R}}

\renewcommand{\P}{\mathbb{P}}
\newcommand{\Q}{\mathbb{Q}}
\newcommand{\var}{\textrm{var}}

\newcommand{\levy}{{L\'evy}}

\newcommand{\Tau}{{\mathcal{T}}}

\title{Tempered Stable Processes with Time Varying Exponential Tails}
\providecommand{\keywords}[1]{\textbf{\textit{Key words:}} #1}
\begin{document}
\begin{center}
\doublespacing
{\Huge
Tempered Stable Processes with Time Varying Exponential Tails
}\\
\vspace{2cm}
{\Large
Young Shin Kim}\\
Associate Professor\\
College of Business, Stony Brook University,\\
100 John S. Toll Drive, Stony Brook, NY 11794, USA\\
Tel: +1 (631) 632-7171 \\
e-mail: aaron.kim@stonybrook.edu \\
~\\
{\Large Kum-Hwan Roh}\footnote{Kum-Hwan Roh gratefully acknowledges the support of Basic Science Research Program through the National Research Foundation of Korea (NRF) grant funded by the Korea government [Grant No. NRF-2017R1D1A3B03036548]. }\\
Associate Professor\\ 
Department of Mathematics, Hannam Univeristy\\
 Deajeon, Korea \\
e-mail: khroh@hnu.kr\\
~\\
{\Large Raphael Douady}\\
Research Professor\\
University of Paris I: Panth{\`e}on-Sorbonne\\
France\\
e-mail: rdouady@gmail.com\\

\end{center}

\clearpage
\maketitle
 
\begin{abstract}
In this paper, we introduce a new time series model having a stochastic exponential tail. This model is constructed based on the Normal Tempered Stable distribution with a time-varying parameter. The model captures the stochastic exponential tail, which generates the volatility smile effect and volatility term structure in option pricing. Moreover, the model describes the time-varying volatility of volatility. We empirically show the stochastic skewness and stochastic kurtosis by applying the model to analyze S\&P 500 index return data. We present the Monte-Carlo simulation technique for the parameter calibration of the model for the S\&P 500 option prices. We can see that the stochastic exponential tail makes the model better to analyze the market option prices by the calibration.
\\
\end{abstract}

\baselineskip=24pt

\doublespacing
\section{Introduction}
The tempered stable process is popularly used as an option pricing model overcoming drawbacks of Black-Scholes model (see \cite{BarndorffNielsenLevendorskii:2001}, \citeauthor{BarndorffNielsenShephard:2001} (\citeyear{BarndorffNielsenShephard:2001}), \cite{CGMY:2002}, and \cite{Kim:2005}) since the class of tempered stable processes are semi-martingale and has exponential tails which are fatter than Gaussian distribution. Moreover, the tempered stable option price model explains the volatility smile and skew effect since its tails are fat and asymmetric. However, the class of tempered stable models' independent and stationary increments fails to capture the stochastic volatility, stochastic skewness, and stochastic kurtosis in the market. In this paper, we construct a new market model that has stochastic exponential tails. Using the stochastic exponential tails, the new model can capture more stochastic properties observed in the market, including stochastic skewness and kurtosis, and volatility of volatility (vol-of-vol).

Managing volatility and vol-of-vol are important issues in portfolio and risk management and derivative pricing. Since they are not directly observed in the market, VIX index (\cite{CBOE:VIX}) and VVIX index (\cite{CBOE:VVIX}) are provided for measuring the volatility and vol-of-vol of U.S. stock market, respectively. Academically, ARCH and GARCH models by \cite{Engle:1982} and \cite{Bollerslev:1986}. The implied volatility extracted from the Black-Scholes model (\cite{BlackScholes:1973}) has been popularly used to observe the volatility.

Applying the ARMA-GARCH model to empirical daily log-returns of a stock or an index, we can see that the residual distribution still has fat-tails and asymmetricity (see \cite{Kim_et_al:2008b,Kim_et_al:2011}). 
In order to capture those the fat-tails and skewness of the residual distribution, ARMA-GARCH model with the standard normal tempered stable innovation distribution (ARMA-GARCH-NTS model) was studied in risk management and portfolio management in many literatures including \cite{Kim:2015}, \cite{Anad_et_al:2016,Anad_et_al:2017} and \cite{KurosakiKim:2018}. The normal tempered stable (NTS) distribution was presented in finance by \cite{BarndorffNielsenLevendorskii:2001} and \cite{BarndorffNielsenShephard:2001} to describe the fat-tail and skewness of asset returns. The standard NTS (stdNTS) distribution is a special case of the NTS distribution with zero mean and unit variance (See \cite{RachevKimBianchiFabozzi:2011a}). 

The volatility clustering and fat-tailed asymmetric distribution have been studied in option pricing in literature. The \levy ~ process model, stochastic volatility model, and GARCH model were introduced to overcome the drawback of Black-Scholes option pricing model. For instance, the \levy~stable model was applied to the option pricing in \cite{Hurst_et_al:1999} and \cite{CarrWu:2003}. The tempered stable option pricing models were discussed in \cite{Boyarchenko_Levendorskii:2002} and \cite{CGMY:2002}. Stochastic volatility model was applied to option pricing in \cite{Heston:1993}, and stochastic volatility model with the \levy~driving process has been studied in \cite{Carr_et_al:2003}. The discrete-time volatility clustering effect was considered for option pricing by taking GARCH model in \cite{Duan:1995}. GARCH option pricing model with non-Gaussian tempered stable innovation was studied by
\cite{Kim_et_al:2010:JBF} and regime-switching tempered stable model were applied to the option pricing in \cite{Kimetal:2012}. The skewness and kurtosis were used in addition to volatility for option pricing in \cite{AbouraMaillard:2016}. Moreover, \levy ~process model with long-range dependence was presented in \cite{KimDanlingStoyan:2019}. 

While the stochastic volatility and volatility clustering were studied, the term structure of vol-of-vol was studied for VIX and VVIX derivatives pricing. For example, the class of \levy~Ornstein Uhlenbeck process is used for modeling vol-of-vol in \cite{MenciaSentana:2013} and the Heston style term structure of vol-of-vol has been presented in \cite{HuangEtAl:2018}, and \cite{BrangerEtAl:2018}. Also, \cite{doi:10.1080/14697688.2017.1412493} considered the Heston style volatility model for the volatility together with the dependence feature between VIX and S\&P 500 index.

In this paper, we will discuss two empirical properties of skewness and excess kurtosis: (1) the residual distribution of S\&P 500 index daily return has negative skewness and large excess kurtosis. Moreover, the absolute value of skewness is increasing then the excess kurtosis is rising together. (2) Skewness and excess kurtosis of S\&P 500 index daily return distribution are not constant but time-varying. 
We will present a new advanced model named the Stochastic Tail NTS (StoT-NTS) model to describe those two properties. In order to construct the model, we take ARMA-GARCH-NTS model and apply a simple time series model to one shape parameter of stdNTS distribution. After constructing the model, a parameter estimation method for the StoT-NTS model will be provided. Using the model, one can capture the time-varying vol-of-vol on stock or index return process.
After the model construction, we apply the model to option pricing. We discuss the Monte-Carlo simulation algorithm for European option pricing on the StoT-NTS model. To verify the model's performance, we calibrate parameters of the model using the S\&P 500 index option prices. As mentioned in the previous paragraph, \cite{AbouraMaillard:2016} considers the skewness and excess kurtosis in option pricing, while we consider a parametric model with stochastic skewness and stochastic kurtosis in option pricing in this paper. The StoT-NTS option pricing model can extract the structure of invisible time-varying vol-of-vol in the market option prices.

The remainder of this paper is organized as follows. The NTS distribution is discussed in Section 2. In Section 3, we present the stochastic properties of skewness and excess kurtosis of the residual distribution for ARMA-GARCH model and empirical study using the S\&P 500 index daily return data. The StoT-NTS model is constructed in this section and shows the model has time-varying vol-of-vol. The option pricing model on the StoT-NTS model is discussed in Section 4. The Monte-Caro algorithm and model calibration are also provided in this section. 
Finally, Section 5 concludes.

\section{Normal Tempered Stable Distribution}
Let $\alpha\in(0,2)$, $\theta, \gamma>0$, and $\mu, \beta\in\R$.
Let $\Tau$ be a positive random variable whose characteristic function $\phi_{\Tau}$ is equal to
\begin{equation}\label{eq:ChF.TSsubordProcess}
\phi_{\Tau}(u) =
\exp\left(-\frac{2\theta^{1-\frac{\alpha}{2}}}{\alpha}\left((\theta-iu)^{\frac{\alpha}{2}}-\theta^{\frac{\alpha}{2}}\right)\right).
\end{equation}
The random variable $\Tau$ is referred to as \emph{Tempered Stable Subordinator}. The \emph{normal tempered stable} (NTS) random variable $X$ with parameters $(\alpha$, $\theta$, $\beta$, $\gamma, \mu)$ is defined as
\begin{equation}\label{eq:multidimStdNTS}
X = \mu-\beta + \beta\Tau + \gamma \sqrt{\Tau}W,
\end{equation}
where $W\sim N(0,1)$ is independent of $\Tau$, 
and we denote $X \sim \textup{NTS}(\alpha$, $\theta$, $\beta$, $\gamma, \mu)$.
The characteristic function (Ch.F) of $X$ is given by
\begin{align*}
&\phi_{NTS}(u) = E[e^{iuX}]\\
&=\exp\left((\mu-\beta)iu-\frac{2\theta^{1-\frac{\alpha}{2}}}{\alpha}
\left(\left(\theta-i\beta u+\frac{\gamma^2 u^2}{2}\right)^{\frac{\alpha}{2}}-\theta^{\frac{\alpha}{2}}\right)\right).
\end{align*}
The first four moments of $X$ are as follows:
\begin{itemize}
\item Mean: $\displaystyle E[X]=\mu$
\item Variance: $\displaystyle \var(X)=\gamma^2+\beta^2\left(\frac{2-\alpha}{2\theta}\right)$
\item skewness: $\displaystyle \textup{S}(X)=
\frac{\beta\,\left(2-\alpha\right)\,\left(6\,{\gamma}^2\,\theta-\alpha\beta^2+4\beta^2\right)}{\sqrt{2\theta}\,{\left(2\,{\gamma}^2\,\theta-\alpha\beta^2+2\beta^2\right)}^{3/2}}
$
\item Excess kurtosis: $\displaystyle \textup{K}(X)=\frac{\left(2-\alpha\right)\,\left({\alpha}^2\beta^4-10\,\alpha\beta^4-12\,\alpha\beta^2\,{\gamma}^2\,\theta+24\beta^4+48\beta^2\,{\gamma}^2\,\theta+12\,{\gamma}^4\,{\theta}^2\right)}{2\,\theta\,{\left(2\,{\gamma}^2\,\theta-\alpha\beta^2+2\beta^2\right)}^2}
$
\end{itemize}
Hence, if $\mu = 0$ and $\gamma=\sqrt{1-\beta^2 \left(\frac{2-\alpha}{2\theta}\right)}$ with $|\beta|<\sqrt{ \frac{2\theta}{2-\alpha}}$ then $\epsilon \sim \textup{NTS}(\alpha$, $\theta$, $\beta$, $\gamma, \mu)$ has zero mean and unit variance.
Put $\beta=B\sqrt{ \frac{2\theta}{2-\alpha}}$ for $B\in (-1,1)$, then $|\beta|<\sqrt{ \frac{2\theta}{2-\alpha}}$ and $\gamma = \sqrt{1-B^2}$.
Then the Ch.F of $\epsilon$ equals to 
\begin{align*}
&\phi_{\epsilon}(u)=E[e^{iu\epsilon}]\\
&
=\exp\left(-iuB\sqrt{ \frac{2\theta}{2-\alpha}}-\frac{2\theta^{1-\frac{\alpha}{2}}}{\alpha}
\left(\left(\theta-iuB\sqrt{ \frac{2\theta}{2-\alpha}}+\frac{ u^2}{2}\left(1-B^2\right)\right)^{\frac{\alpha}{2}}-\theta^{\frac{\alpha}{2}}\right)\right)
\end{align*}
In this case $\epsilon$ is referred to as the \textit{standard NTS} random variable with parameters $(\alpha,\theta; B)$, and we denote $\epsilon\sim \textup{stdNTS} (\alpha,\theta; B)$.\footnote{
The standard NTS distribution is defined by the NTS distribution with $\mu = 0$ and $\gamma=\sqrt{1-\beta^2 \left(\frac{2-\alpha}{2\theta}\right)}$ under the condition that $|\beta|<\sqrt{ \frac{2\theta}{2-\alpha}}$ and denoted to $\textup{stdNTS}(\alpha, \theta, \beta)$, in many literature including \cite{KimKim:2018}, \cite{Anad_et_al:2016}, \cite{Anad_et_al:2017}, and \cite{KIM2015512}. In this paper, we change the parameterization for the convenience.} The Ch.F is denoted by $\phi_{stdNTS}(u;\alpha,\theta; B)=\phi_{\epsilon}(u)$.
For $\epsilon,$ we have
\begin{equation}\label{eq:SkewOfEps}
S(\epsilon) = \sqrt{\frac{2-\alpha}{2\theta}}B\left(3(1-B^2)+\frac{4-\alpha}{2-\alpha}B^2\right)
\end{equation}
and
\begin{equation}\label{eq:KurtOfEps}
K(\epsilon) = \frac{
	\left(
		2-\alpha
	\right)}
{2\theta}
	\left(
		(\alpha-4)(\alpha-6)
		\left(\frac{B^2}{2-\alpha}\right)^2
		+\left(
			(24-6\alpha)\left(\frac{B^2}{2-\alpha}\right)+3(1-B^2)
		\right)
		(1-B^2)
	\right).
\end{equation}

\begin{figure}
\begin{center}
\includegraphics[width=12cm]{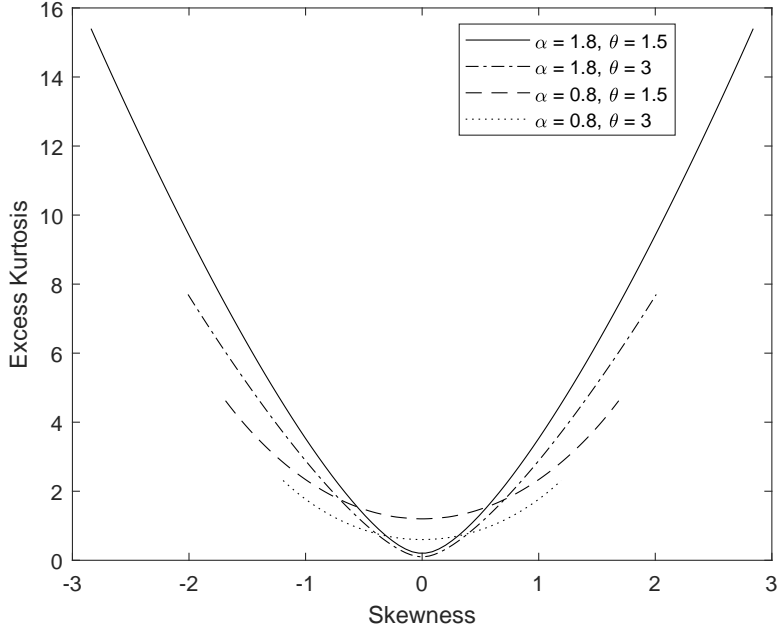}
\caption{\label{Fig:S2K}Graph of skewness to Excess kurtosis for $\epsilon\sim \textup{stdNTS} (\alpha, \theta; B)$ with $(\alpha,\theta)\in\{(1.8,1.5)$, $(1.8,3)$, $(0.8,1.5)$, $(0.8,3)\}$ and $B\in [-1,1]$.}
\end{center}
\end{figure}

Suppose that $\alpha$ and $\theta$ are fixed then we have a function
\[
B\mapsto (\textup{S}(\epsilon), \textup{K}(\epsilon)), ~~~ \theta>0.
\]
We can easily check the following facts:
\begin{itemize}
\item if $B=0$ 
\[
\textup{S}(\epsilon)=0 ~~~\text{ and }~~~ \textup{K}(\epsilon)= \frac{3}{2\theta}(2-\alpha).
\]
\item if $B = \pm 1$ then $\gamma = \sqrt{1-B^2} = 0$, and hence
\[
S(\epsilon) = \frac{\pm(4-\alpha)}{\sqrt{2\theta(2-\alpha)}} ~~~\text{ and }~~~
K(\epsilon) = \frac{(\alpha-4)(\alpha-6)}{2\theta(2-\alpha)}.
\]
\end{itemize}
For example,
\begin{itemize}
\item if $\alpha = 1.8$ and $\theta = 1.5$, then $\textup{S}(\epsilon)\in[-2.8402, 2.8402]$ and $\textup{K}(\epsilon) \in[0.2,15.4]$.
\item if $\alpha = 0.8$ and $\theta = 3$, then $\textup{S}(\epsilon)\in[-1.1925, 1.1925]$ and $\textup{K}(\epsilon) \in[0.6,2.3111]$.
\end{itemize}
Other example cases of the function are presented in the Figure \ref{Fig:S2K}. The points of $(\textup{S}(\epsilon), \textup{K}(\epsilon))$ are smoothly connected parabolic curve for $B\in [-1,1]$.

\section{ARMA-GARCH-NTS Model with Stochastic Parameter $B$}
Taking the ARMA(1,1)-GARCH(1,1) model as
\begin{equation*}
\begin{cases}
y_{t+1}=c+ay_{t}+b\sigma_{t}\epsilon_{t} + \sigma_{t+1}\epsilon_{t+1}\\
\sigma_{t+1}^2 = \kappa + \xi \sigma_{t}^2\epsilon_{t}^2 + \zeta\sigma_{t}^2
\end{cases}, ~~~
\end{equation*}
we assume that $\epsilon_t\sim \textup{stdNTS}(\alpha,\theta; B)$. Then we obtain the ARMA-GARCH-NTS model. Suppose that the parameter $\alpha$ and $\theta$ are fixed real numbers, and parameter $B$ is replaced to a random variable, then we obtain a new time series model. In this paper, we assume that 
\begin{itemize}
\item $(\epsilon_t)_{t=1,2,\cdots}$ is not i.i.d, but $\epsilon_{t|t-1}\sim\textup{stdNTS}(\alpha, \theta; B_t)$,
\item and $(B_t)_{t=1,2,\cdots}$ is given by a  ARIMA(1,1,0) model as follows:
\begin{align*}
B_{t+1} &= B_{t} + \varDelta B_{t+1} \\
\varDelta B_{t+1} &= a_0 + a_1 \varDelta B_{t} + \sigma_Z Z_{t+1}, 
\end{align*}
where $a_0, a_1\in\R$, $|a_1|<1$, $\sigma_Z>0$, and $(Z_t)_{t=1,2,\cdots}$ is i.i.d with $Z_t\sim N(0,1)$.
\end{itemize}
This time series model is referred to as the \textit{Stochastic Tails ARMA-GARCH-NTS} model or shortly the \textit{StoT-NTS} model. 

Note that, the conditional skewness of  $\sigma_{t}\epsilon_{t}$ is given as
\begin{align*}
S(\sigma_{t}\epsilon_{t}|\mathcal F_{t-1})&=S(\epsilon_{t}|\mathcal F_{t-1})
\\
&=\sqrt{\frac{2-\alpha}{2\theta}}B_t\left(3(1-B_t^2)+\frac{4-\alpha}{2-\alpha}B_t^2\right)
\end{align*}
by \eqref{eq:SkewOfEps}. Moreover, the conditional variance of variance for $\sigma_{t+1}\epsilon_{t+1}$ is 
\begin{align*}
\var(\var(\sigma_{t+1}\epsilon_{t+1}|\mathcal{F}_t)|\mathcal{F}_{t-1}) 
&= \xi^2(\sigma_{t|\mathcal F_{t-1}})^4 E[\epsilon_{t|\mathcal F_{t-1}}^4] 
\\&
= \xi^2(\kappa + \xi \sigma_{t-1}^2\epsilon_{t-1}^2 + \zeta\sigma_{t-1}^2)^2 K(\epsilon_{t}|\mathcal F_{t-1}).
\end{align*}
Since $\epsilon_{t|{t-1}}\sim \textup{stdNTS}(\alpha,\theta; B_t)$, we obtain 
\begin{align*}
&\var(\var(\sigma_{t+1}\epsilon_{t+1}|\mathcal{F}_t)|\mathcal{F}_{t-1}) \\
& = 
\frac{
	\left(
		2-\alpha
	\right)}
{2\theta}
\xi^2(\kappa + \xi \sigma_{t-1}^2\epsilon_{t-1}^2 + \zeta\sigma_{t-1}^2)^2 
\\&
\times	\left(
		(\alpha-4)(\alpha-6)
		\left(\frac{B_t^2}{2-\alpha}\right)^2
		+\left(
			(24-6\alpha)\left(\frac{B_t^2}{2-\alpha}\right)+3(1-B_t^2)
		\right)
		(1-B_t^2)
	\right),
\end{align*}
by \eqref{eq:KurtOfEps}. Hence,
the StoT-NTS process captures the time varying skewness and time varying vol-of-vol for the random variable $B_t$.

\subsection{\label{Sec:ARMA_GARCH_Est}ARMA-GARCH parameter estimation}
\begin{figure}
\begin{center}
\includegraphics[width=8cm]{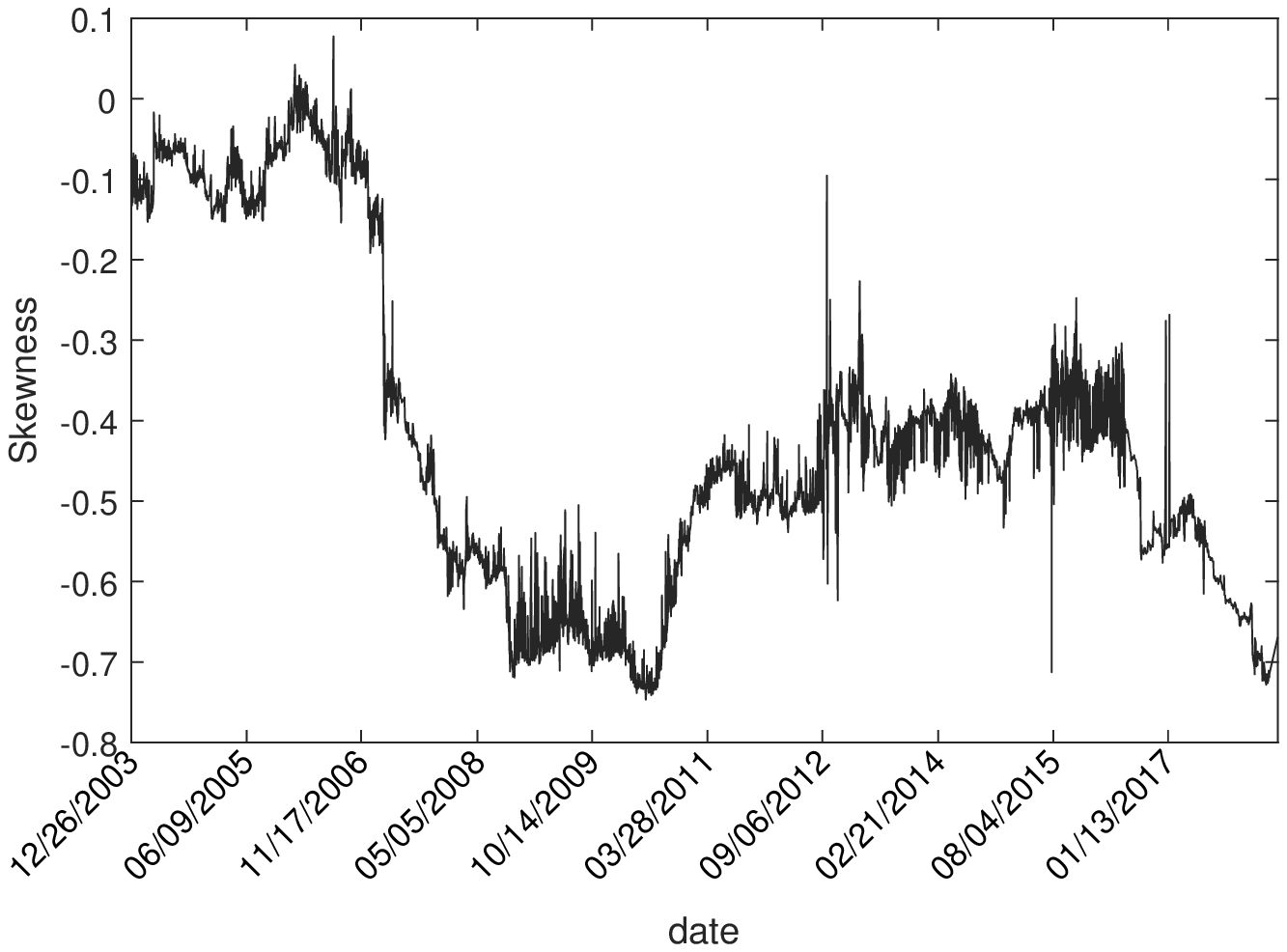}
\includegraphics[width=8cm]{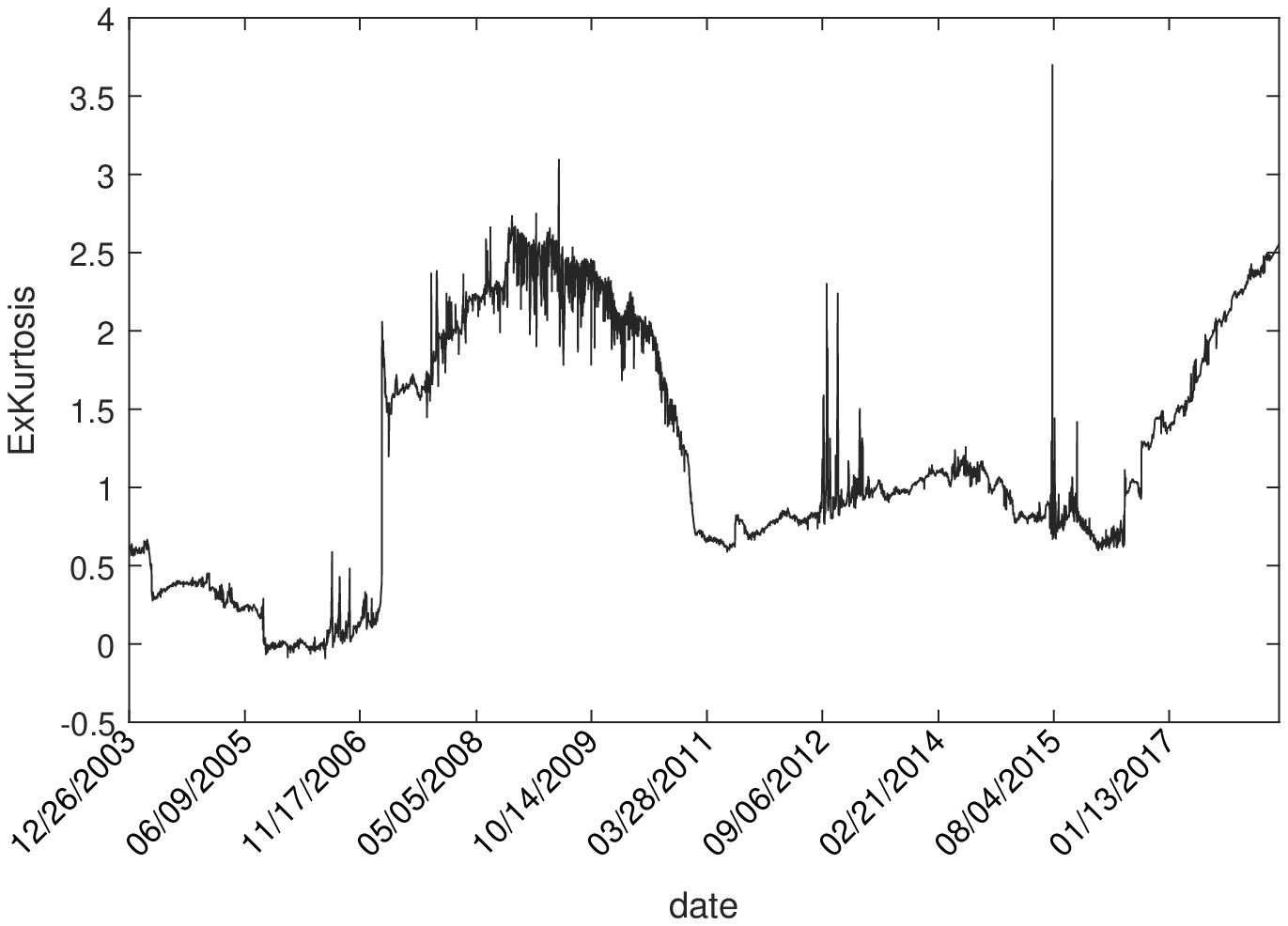}
\caption{\label{Fig:SkKurTimeseries}Time series of empirical skewness and excess kurtosis  for each residual sets $R_1, R_2, \cdots, R_{3607}$. }
\end{center}
\end{figure}

\begin{figure}
\begin{center}
\includegraphics[width=10cm]{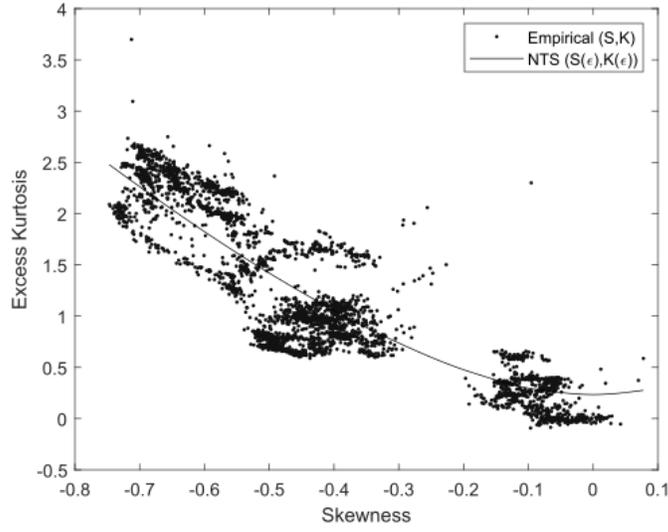}
\caption{\label{Fig:EmpS2K}Dots are empirical excess kurtosis values and their corresponding empirical skewness values for set of residuals $R_t\in\{R_1, R_2, \cdots, R_{3607}\}$. The solid curve is the curve of excess kurtosis and skewness for $\epsilon\sim\textup{stdNTS}(\alpha,\theta; B)$ with estimated parameters $\alpha = 1.8043$ and $\theta = 1.2544$.}
\end{center}
\end{figure}

We estimate model parameters using S\&P 500 index daily log-return data.
ARMA(1,1)-GARCH(1,1) parameters are estimated for every 3,607 working days between December 26, 2003 to June 1, 2018. In each estimation, we use 1,000 historical log-returns by the current day.
For example,
\begin{itemize}
\item at December 26, 2003, we estimate those parameters using 1,000 daily log-returns from January 4, 2000 to December 26, 2003,
\item at June 1, 2018, we estimate those parameters using 1,000 daily log returns from May 7, 2014 to June 1, 2018.
\end{itemize}
Then we obtain 3,607 residual sets. Each residual set contains 1,000 elements extracted from the estimation.
Let $R_1, R_2, \cdots, R_{3607}$ be those residual sets. For instance, $R_1$ is the residual set extracted from the ARMA(1,1)-GARCH(1,1) estimation at December 26, 2003, and $R_{3607}$ is the residual set extracted from the estimation at June 1, 2018.
We calculate  empirical skewness $\textup{S}(R_t)$ and empirical excess kurtosis $\textup{K}(R_t)$ for $R_t\in\{R_1,$ $R_2,$ $\cdots,$ $R_{3706}\}$. Then we obtain the skewness time series $(\textup{S}(R_t))_{t=1,2,\cdots, 3607}$ and excess kurtosis time series $(\textup{K}(R_t))_{t=1,2,\cdots, 3607}$, which are presented in Figure \ref{Fig:SkKurTimeseries}. Moreover, we plot pairs of excess kurtosis and skewness $(S(R_t), K(R_t))$ for $t\in\{1,2,\cdots, 3607\}$ as Figure \ref{Fig:EmpS2K}. 
We found that, negative skewness leads large excess kurtosis, and small excess kurtosis follows zero skewness.

\subsection{ Fit parameters $\alpha$ and $\theta$}
We fit $\alpha$ and $\theta$ of the stdNTS process as follows:
\begin{itemize}
\item Select one $\alpha\in(0,2)$ and one $\theta>0$. Let $M_{\alpha,\theta}=\{(\textup{S}(\epsilon),\textup{K}(\epsilon))\,|\, \epsilon\sim \textup{stdNTS}( \alpha,\theta;B) $ for $B\in[-1,1]\}$.
\item Applying interpolation for $M_{\alpha, \theta}$, we define a function $f_{\alpha, \theta}$ from skewness to excess kurtosis. That is,  
\[
f_{\alpha, \theta}(\textup{S}(\epsilon)) =  \textup{K}(\epsilon)
\text{ for } (\textup{S}(\epsilon),\textup{K}(\epsilon))\in M_{\alpha,\theta}.
\] 
\item Find optimal $(\alpha^*, \theta^*)$ minimize the square error for the empirical data as
\[
(\alpha^*, \theta^*) = \textup{arg} \min_{(\alpha, \theta)} \sum_{t=1}^{T} \left[f_{\alpha, \theta}(\textup{S}(R_t))-\textup{K}(R_t))\right]^2/T
\]
where $T=3607$.
\end{itemize}
By the fitting method, we obtained $(\alpha^*, \theta^*) = (1.8043, 1.2544)$, and the solid curve in Figure \ref{Fig:EmpS2K} is the function $f_{\alpha^*, \theta^*}$.

\subsection{Fit parameters for the time series $(B_t)_{t\ge0}$}
\begin{figure}
\begin{center}
\includegraphics[width=12cm]{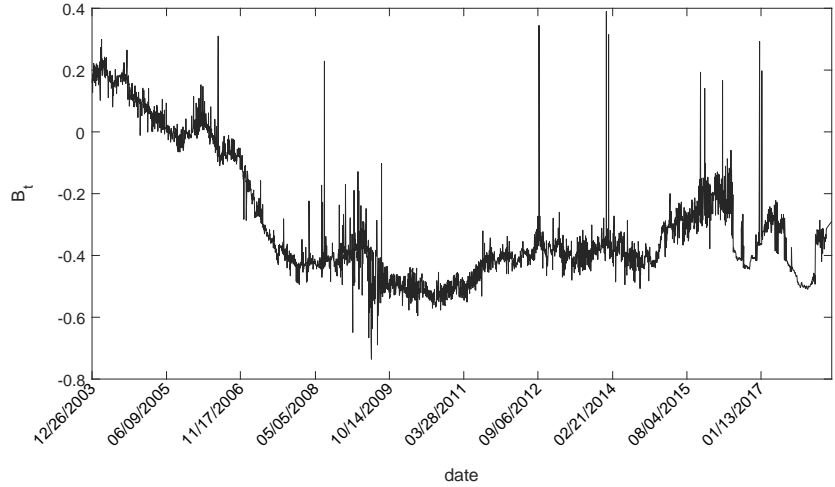}
\caption{\label{Fig:BTimeSeries}The time series of the estimated $B_t$ for each residual set in $\{R_1, R_2,  \cdots, R_{3607}\}$}
\end{center}
\end{figure}

\begin{figure}
\includegraphics[width=16cm]{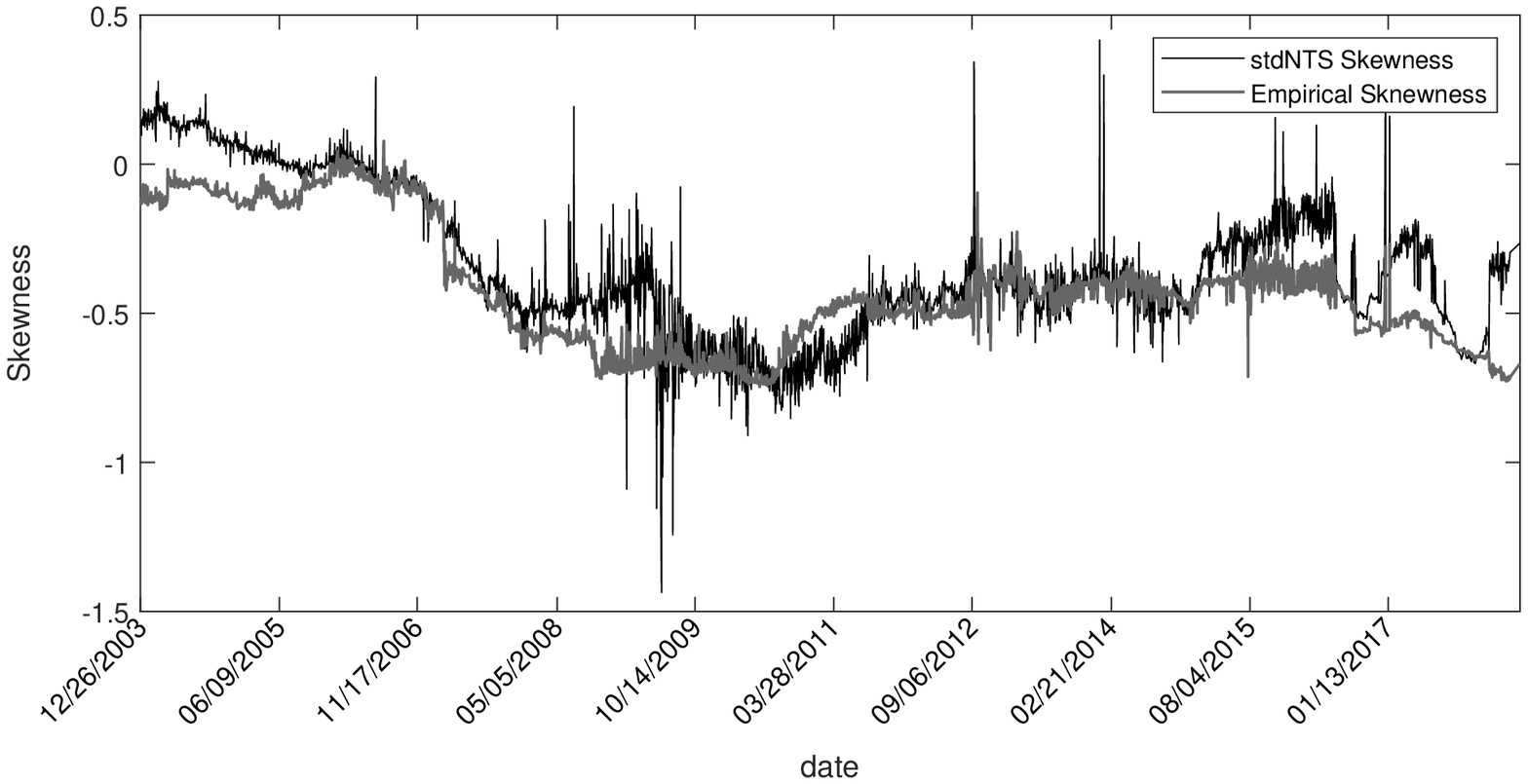}\\
\includegraphics[width=16cm]{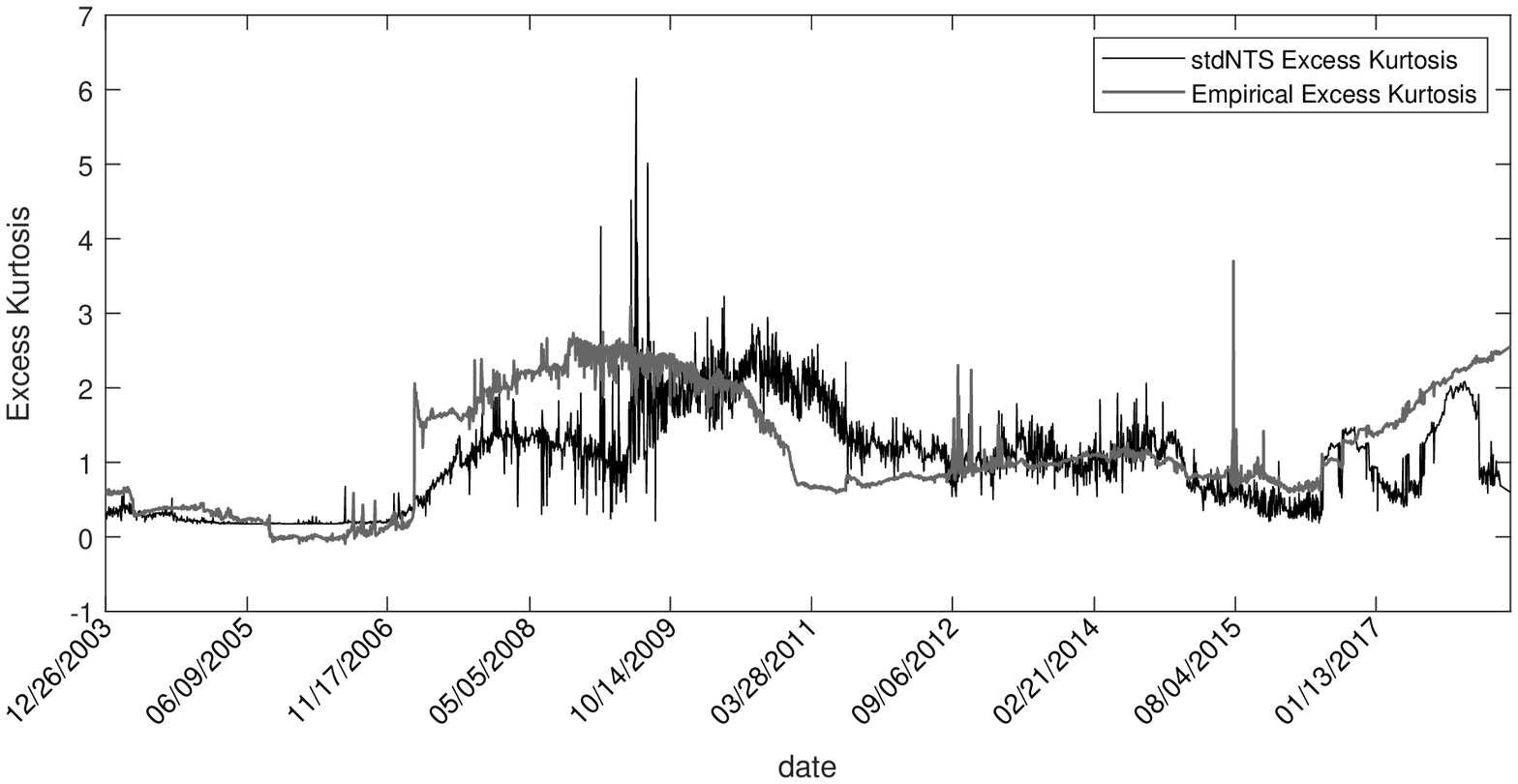}
\caption{\label{Fig:EmpKurVsntsKur}Gray curses are time series of empirical skewness and excess kurtosis and black curses are time series of stdNTS skewness and excess kurtosis for each residual set in $\{R_1, R_2,  \cdots, R_{3607}\}$. }
\end{figure}

We fix  parameters $\alpha = 1.8043$ and $\theta = 1.2544$, and fit parameter $B$ of stdNTS to the daily residual set $R_t$ for $t\in\{1,2,\cdots, 3607\}$. In this parameter fit, we find the empirical cdf $F^{emp}_t$ using KS-density for $R_t$ and find $B$ using the least square curve fit as
\[
B_t = \textup{arg} \min_{B} \sum_{x_k\in R_t} (F(x_k; \alpha, \theta, B)-F_t^{emp}(x_k))^2
\]
where $F(x;\alpha, \theta, B)$ is the CDF of stdNTS$(\alpha, \theta; B)$.  
Figure \ref{Fig:BTimeSeries} presents the time series of the estimated $B_t$ for daily residual $R_t$ with $t\in\{1,2,\cdots, 3607\}$. Figure \ref{Fig:EmpKurVsntsKur} has two plates. The upper plate exhibits the empirical skewness time series and the skewness time series of stdNTS($\alpha$, $\theta$; $B_t$), and the bottom plate provides the empirical excess kurtosis time series and the excess kurtosis of stdNTS($\alpha$, $\theta$; $B_t$), where 
$\alpha = 1.8043$, $\theta = 1.2544$ and $B_t$ in Figure \ref{Fig:BTimeSeries}.

We apply the ARIMA(1,1,0) model to the time series $(B_t)_{t\ge0}$ given in Figure \ref{Fig:BTimeSeries} as
\[
\varDelta B_{t+1} = c_B + a_B\varDelta B_t + \sigma_B Z,
\]
where $\varDelta B_{t+1} = B_{t+1}-B_t$.
We obtain the ARIMA(1,1,0) parameters as (a) of Table \ref{Table:AR1Paramest}. The constant $c_B$ is not significant at $5\%$ significant level, and hence we can set $c_B=0$. The AR parameter $a_B$ and the variance $\sigma_B^2$ are significant. Set the constant $c_B=0$ , and re-estimate ARIMA(1,1,0) we obtain (b) of Table \ref{Table:AR1Paramest} which is similar to (a). We observe the negative AR parameter, that is, $\varDelta B_t$ is mean reverting.

\begin{table}
\caption{\label{Table:AR1Paramest}AR(1) Parameter Estimation}
\begin{center}
(a)\\
\begin{tabular}{ccccc}
\hline
	& Value & Standard Error & t-statistic & $p$-value\\
\hline
$c_B$   & $-0.00018989$   &       $0.00095468$   &       $-0.1989$   &       $0.84234$\\
$a_B$   &       $-0.47936$   &       $0.003392$   &       $-141.32$   &       $0$\\
$\sigma_B^2$ &      $0.0028331$   &       $1.4057\cdot 10^{-5}$   &       $201.55$   &       $ 0$\\
\hline
\end{tabular}
\end{center}
\begin{center}
(b)\\
\begin{tabular}{ccccc}
\hline
	& Value & Standard Error & t-statistic & $p$-value\\
\hline
$a_B$   &       $-0.47935$   &       $0.0033899$   &       $-141.4$   &       $0$\\
$\sigma_B^2$ &      $0.0028331$   &       $1.3057\cdot 10^{-5}$   &       $216.99$   &       $ 0$\\
\hline
\end{tabular}
\end{center}
\end{table}

\section{Option Pricing on the StoT-NTS Model}

Let $(S_t)_{t\in\{0,1,\cdots, T^*\}}$ be the underlying asset price process and $(y_t)_{t\in\{0,1,2,\cdots T^*\}}$ be the underlying asset log return process ($y_0=0$) with $y_t = \log(S_t/S_{t-1})$ where $T^*<\infty$ in the time horizon.
Under the physical measure $\P=\bigoplus_{t=1}^{T^*}\mathcal P_t$, $(y_t)_{t\in\{0,1,2,\cdots T^*\}}$ is supposed to follow the StoT-NTS model:
\begin{equation*}
\begin{cases}
y_{t+1}= \mu_{t+1} + \sigma_{t+1}\epsilon_{t+1|t}\\
\mu_{t+1} = c+ay_{t}+b\sigma_{t}\epsilon_{t|t-1}\\
\sigma_{t+1}^2 = \kappa + \xi \sigma_{t}^2\epsilon_{t|t-1}^2 + \zeta\sigma_{t}^2
\end{cases}
\end{equation*}
where $\epsilon_{t+1|t}\sim \textup{stdNTS}(\alpha, \theta ;B_{t+1})$, 
with
\[
B_{t+1} = B_{t} + a_B \varDelta B_{t} + \sigma_B Z_{t+1}, ~~~ Z_{t+1}\sim N(0,1)
\]
for $t\in\{0, 1,2,\cdots T^*\}$.
Here, $\Tau$, $W$ and $(Z_t)_{t\in\{1,2,\cdots, T^*\}}$ are mutually independent, and $\epsilon_{0}$ and $\varDelta B_0$ are real constants.
Let $(r_t)_{t\in\{1,2,\cdots,T^*\}}$ be sequence of the daily risk-free rate of return.
There is risk-neutral measure $\Q=\bigoplus_{t=1}^{T^*}\mathcal Q_t$ such that
\begin{itemize}
\item $
\eta_{t+1|t} = \lambda_{t+1} + \epsilon_{t+t|t}
$

where $\lambda_{t+1} = \frac{\mu_{t+1}-r_{t+1}+w_{t+1}}{\sigma_{t+1}}$ with $\omega_{t+1}=\log\left(\phi_{stdNTS(\alpha, \theta, B_t)}(-i\sigma_{t+1})\right)$
\item $\eta_{t+1|t}\sim \textup{stdNTS}( \alpha, \theta; B_t)$ under the measure $\Q$ with
\[
B_{t+1} = B_{t} + a_B \varDelta B_{t} + \sigma_B Z_{t+1}, ~~~ Z_{t+1}\sim N(0,1), t = 0,1,
\cdots, T^*.
\]
\end{itemize}
Hence we have
\begin{equation*}
\begin{cases}
y_{t+1}=r_{t+1} - \omega_{t+1} + \sigma_{t+1}\eta_{t+1|t}\\
\sigma_{t+1}^2 = \kappa + \xi \sigma_{t}^2(\eta_{t|t-1}-\lambda_t)^2 + \zeta\sigma_{t}^2
\end{cases}
\end{equation*}
which is the risk-neutral price process.

Under the risk-neutral measure $\Q$, the underlying asset price is
$S_t = S_0e^{\sum_{j=0}^t y_j}$ for $t\in\{0,1,2,\cdots, T^*\}$. The European option with a payoff function $H(S(T))$ at the maturity $T$ with $t\le T\le T^*$ is given by
\[
E_\Q\left[e^{-r(T-t)}H(S(T))|\mathcal F_t\right]=E_\Q\left[e^{-r(T-t)}H(S_te^{\sum_{j=t}^T y_j})|\mathcal F_t\right].
\]
For example, European vanilla call and put price with strike price $K$ and time to maturity $T$ at time $t=0$ are 
\[
C(K,T) = E_\Q\left[e^{-rT}\max\{S_0e^{\sum_{j=0}^T y_j}-K,0\}\right]
\]
and
\[
P(K,T) = E_\Q\left[e^{-rT}\max\{K-S_0e^{\sum_{j=0}^T y_j},0\}\right]
\]
respectively.

\subsection{Monte-Carlo Simulation and Calibration}
Assume $r_t=r$ and $\lambda_t=\lambda$ constant, to simplify the model.
Let $M$ be the number of scenarios and $T$ be the time to maturity as a positive integer value, say days to maturity. 
\begin{itemize}

\item Step 1\\
Generate a set of uniform random numbers between 0 and 1 ($\mathcal U(0,1)$), and
two sets of independent standard normal ($N(0,1)$) random numbers
\[
u_{m,n}\sim \mathcal U(0,1), x_{m,n} \sim N(0,1) \text{ and } z_{m,n} \sim N(0,1)
\]
for $m = \{1,2,\cdots, M\}$ and $n = \{1,2,\cdots, T\}$.
\item Step 2\\
Generate the tempered stable subordinator by inverse transform algorithm, as
\[
\tau_{m,n} = F_{TS(\alpha,\theta)}^{inv} (u_{m,n})
\]
where $F_{TS(\alpha,\theta)}^{inv}$ is the inverse CDF of tempered stable subordinator with parameter $(\alpha, \theta)$.
\item Step 3\\
Simulate $(B_t)_{0\le t\le T}$ as $(B_{m,n})_{m\in\{1,2,\cdots, M\}, n\in\{1,2, \cdots, T\}}$, where
\[
B_{m,n} = B_{m,n-1} + a_B (B_{m,n-1}-B_{m,n-2}) + \sigma_B z_{m,n}
\]
and $B_{m,0}$ is $B_0$ value at current time, and $B_{m,1}-B_{m,0}=0$.
\item Step 4\\
Using \eqref{eq:multidimStdNTS}, we simulate random number $(\eta_t)_{0\le t\le T}$ as $(\eta_{m,n})_{m\in\{1,2,\cdots, M\}, n\in\{1,2, \cdots, T\}}$, where
\[
\eta_{m,n} = B_{m,n}\sqrt{ \frac{2\theta}{2-\alpha}}(\tau_{m,n}-1) +  x_{m,n}\sqrt{(1-B_{m,n}^2)\tau_{m,n}}.
\]
\item Step 5\\
Generate $\sigma_t$,
\[
\sigma_{m,n} = \sqrt{\kappa+\xi\sigma_{m,n-1}^2(\eta_{m,n-1}-\lambda)^2+\zeta\sigma_{m,n-1}^2}
\]
$\sigma_{m,0}$ is the currently observed volatility,
and generate $y_t$ using $\sigma_t$ by GARCH option pricing model as follows
\[
y_{m,n} = r-w_{m,n}+\sigma_{m,n} \eta_{m,n},
\]
where
 $\omega_{m,n}=\log\left(\phi_{stdNTS(\alpha, \theta, B_{m,n})}(-i\sigma_{m,n})\right)$
\item Step 6\\
The price process is obtained by 
\[
S_{m,n} = S_0 \exp\left(\sum_{j = 1}^ny_{m,j}\right),
\]
for $m = \{1,2,\cdots, M\}$ and  $n = \{1,2,\cdots, T\}$.
\end{itemize}

\begin{figure}
\includegraphics[width = 5cm]{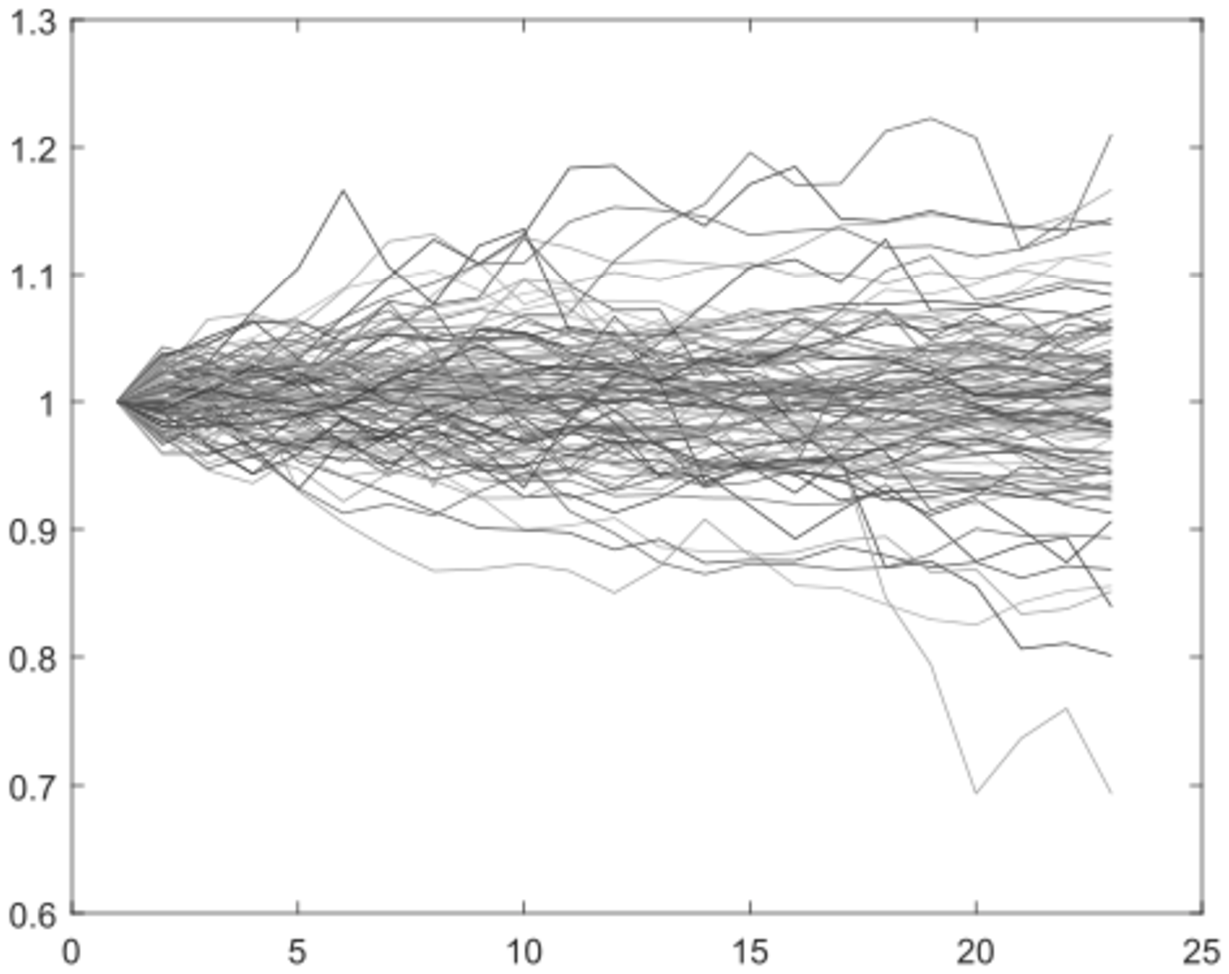}
\includegraphics[width = 5cm]{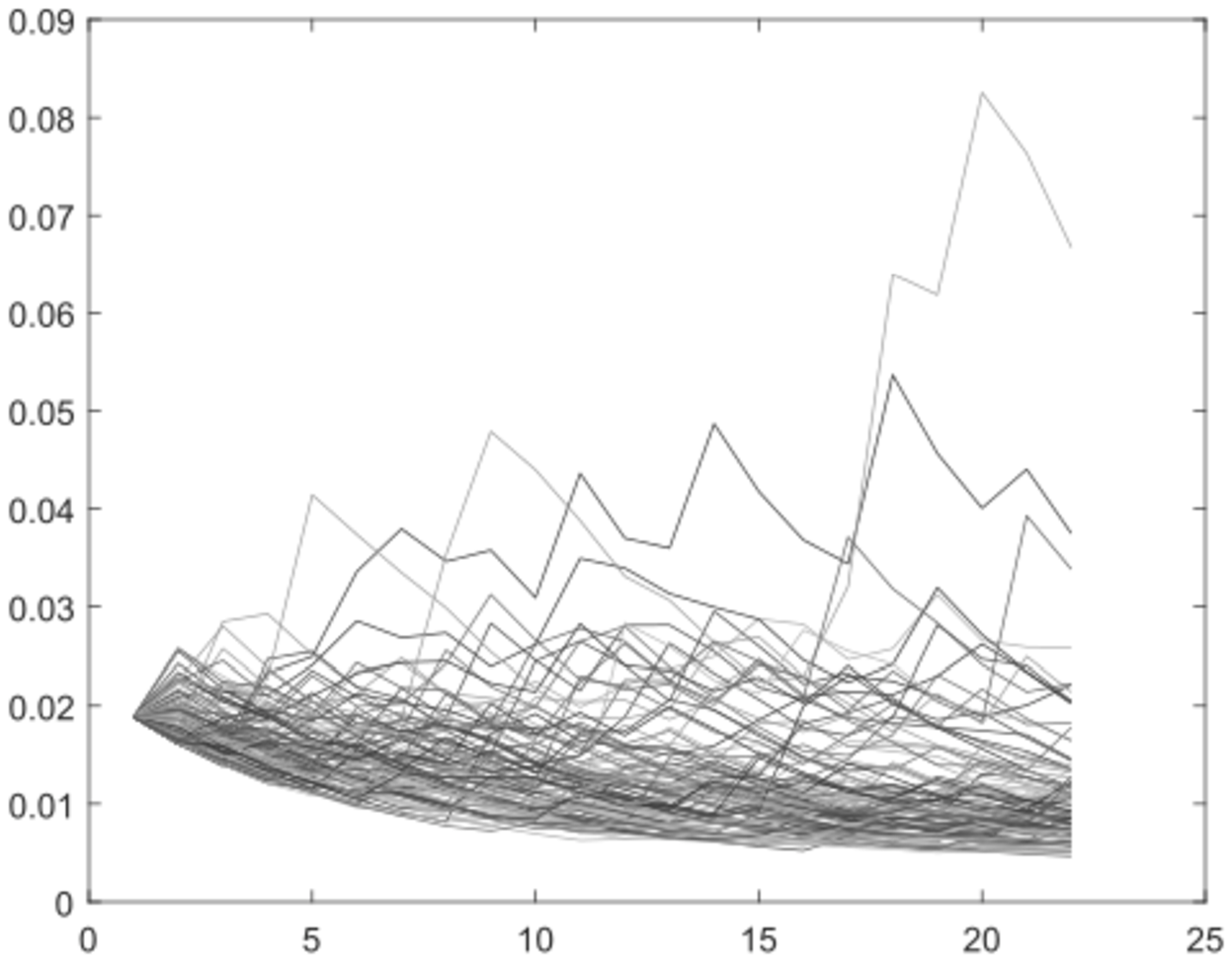}
\includegraphics[width = 5cm]{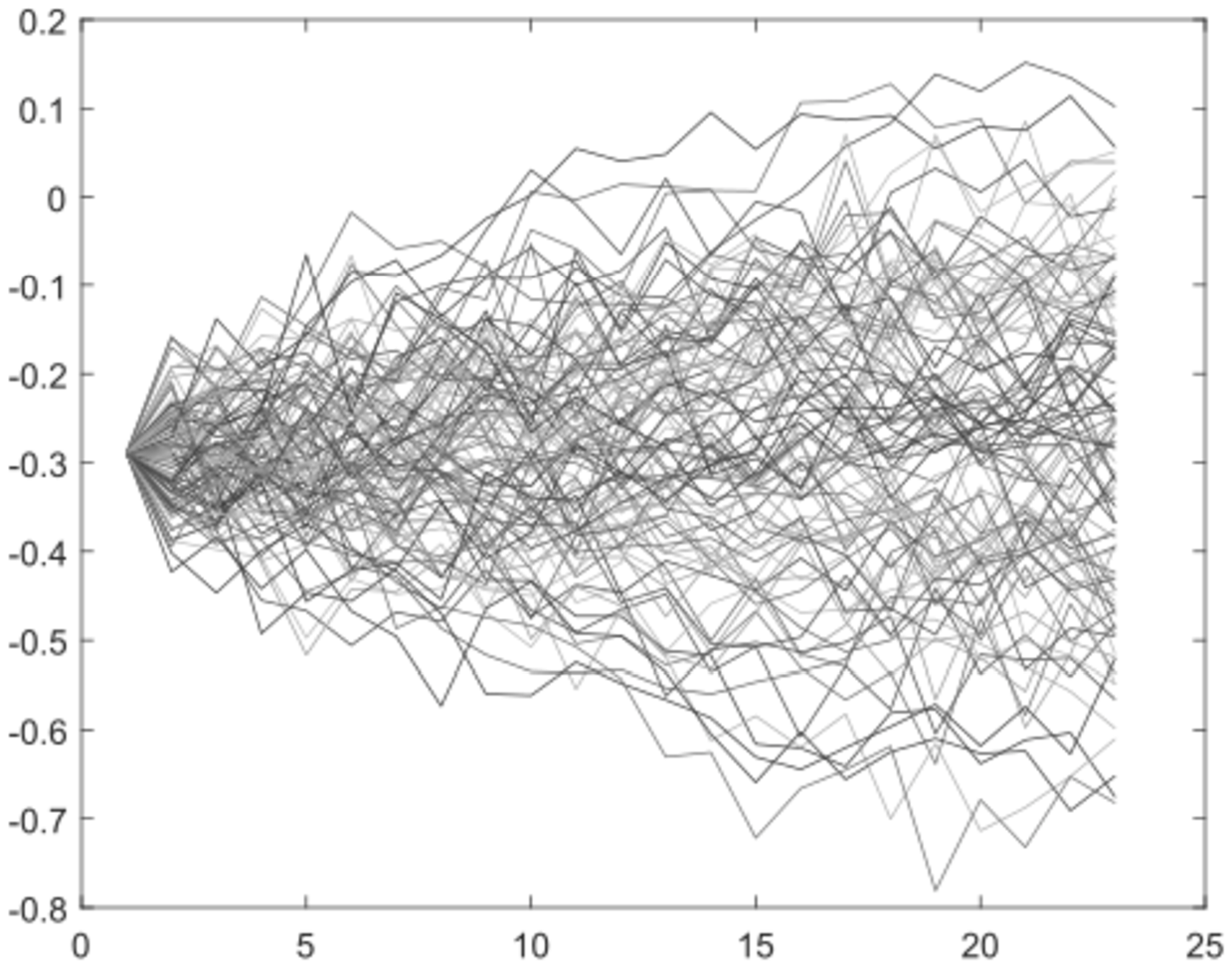}
\caption{\label{Fig_SamplePaths}Simulated $(S_{m,n})$, $(\sigma_{m,n})$, and $(B_{m,n})$ from left.} 
\end{figure}

For example, let GARCH parameters be $\kappa=4.4115\cdot 10^{-6}$, $\xi = 0.2289$, $\zeta = 0.7177$, ARIMA(1,1,0) parameters for $(B_t)$ be $a_B = -0.4793$, $\sigma_B = 0.0532$ and $B_0=-0.2895$, and tempered stable subordinator parameters be $\alpha = 1.8245$ and $\theta = 1.5063$. Set initial values of return, residual and volatility as $y_0 = 0.0373$, $\epsilon_0=3.6851$ and $\sigma_0 = 0.0096$, respectively. Assume that $r = (1/250)\%$, $d = 0$, and $\lambda = 0$, and generate the sample path using the algorithm for $T = 22$ and $M = 100$. Then we obtain the sample path of $(S_{m,n})$ for $S_0=1$, $(\sigma_{m,n})$, and $(B_{m,n})$ as Figure \ref{Fig_SamplePaths}.

The European call option and put option prices with the strike price $K$ and time to maturity $T$ can calculated by the simulated price process as follows:
\begin{align*}
C(K,T) = \frac{e^{-rT}}{M}  \sum_{m=1}^M \max\{S_{m, T}-K, 0\} \\
P(K,T) = \frac{e^{-rT}}{M}  \sum_{m=1}^M \max\{K-S_{m, T}, 0\} .
\end{align*}

\subsection{Calibration}
We calibrate the StoT-NTS parameters using the S\&P 500 index call and put data for the second Wednesday of each month from January 2016 to December 2017. For each calibration date, we use the GARCH parameters estimated in Section \ref{Sec:ARMA_GARCH_Est}.
Table \ref{Table:GARCHparameters} provides those GARCH parameters, and volatility ($\sigma_0$) and and residual ($\epsilon_0$) observed of each date. Daily risk free rates of return and daily continuous dividend rates are also presented in base-point (bp) unit.
We calculate other parameters ($\alpha$, $\theta$, $a_B$, $\sigma_B$, $B_0$, and $\lambda$) for call and put out-of-the-money (OTM) option prices. We generate one set of uniform random numbers and two sets of standard normal random numbers in Step 1 for $M=10,000$ and $T = 90$, and fix them. After then find parameters such that minimize the mean square errors between the model price and the market prices: 
\[
\min_\Theta \left(\sum_{K_n<S_0, ~ T_n<90} (P(K_n,T_n)-P_{market}(K_n,T_n))^2 +\sum_{K_n>S_0, ~ T_n<90} (C(K_n,T_n)-C_{market}(K_n,T_n))^2\right), 
\]
for $\Theta$ = $(\alpha$, $\theta$, $a_B$, $\sigma_B$, $B_0$, $\lambda)$,
where $S_0$ is S\&P 500 index price of the given Wednesday and $P_{market}(K_n,T_n)$ and $C_{market}(K_n,T_n)$ are mid-price of observed bid and ask prices for the call and put on the given day with strike price $K_n$ and time to maturity $T_n$. 

\begin{sidewaysfigure}
\begin{center}
\includegraphics[width = 20cm]{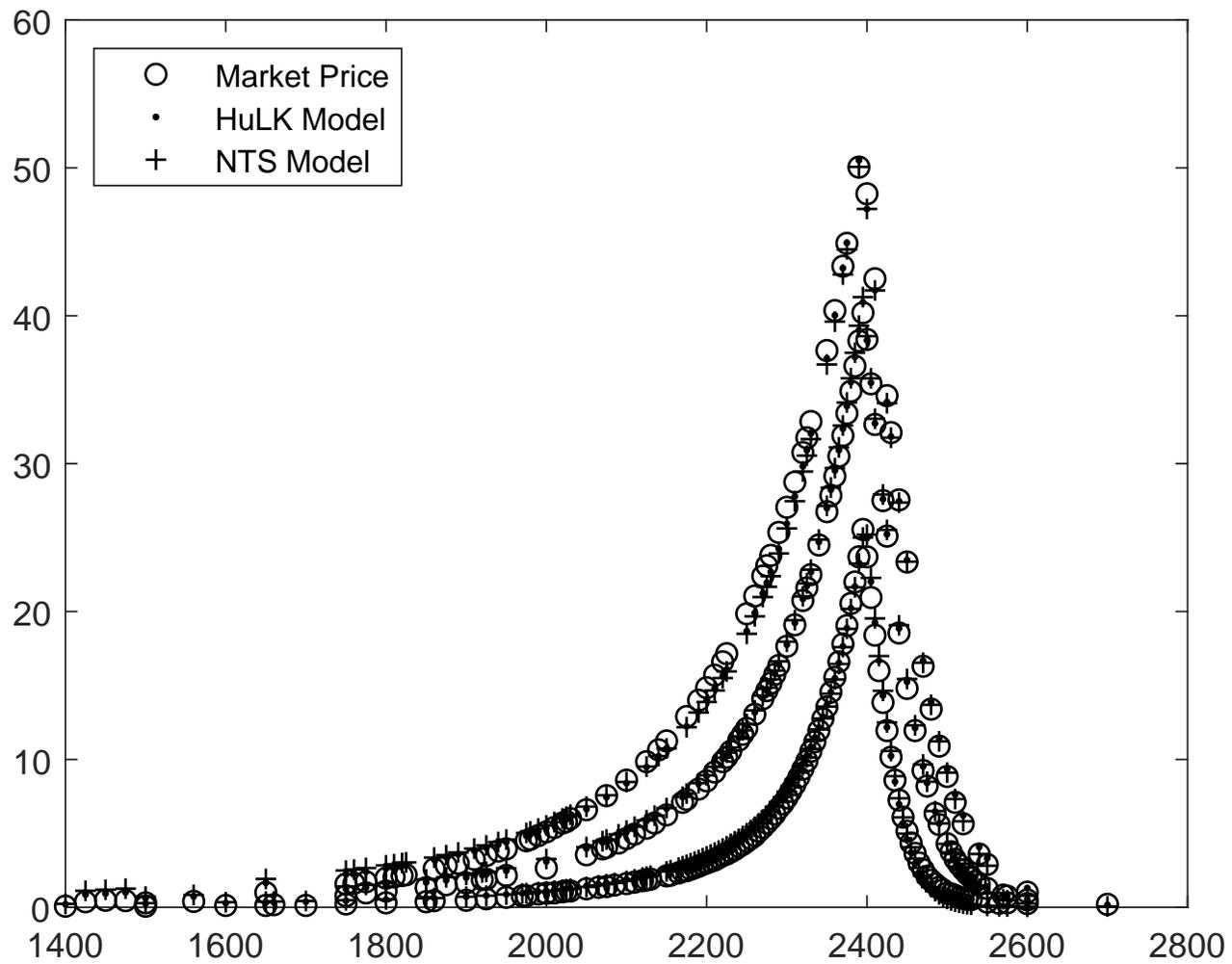}
\caption{\label{fig:OTMPrices} OTM call and put option prices and calibration result for 5/10/2017. Circles ($\circ$) are market prices of calls and puts. Dot ($\cdot$) and plus ($+$) marks are calibrated the StoT-NTS model and the GARCH-NTS model prices, respectively. }
\end{center}
\end{sidewaysfigure}

For example, Figure \ref{fig:OTMPrices} exhibits the market prices and calibrated the StoT-NTS model prices for the OTM call and puts on 5/10/2017. The calibrated GARCH-NTS model prices obtained by the simulation method are presented in the figure as a benchmark model.
The GARCH-NTS model is option pricing model as
\begin{equation*}
\begin{cases}
y_{t+1}=r_{t+1} - \omega_{t+1} + \sigma_{t+1}\eta_{t+1|t}\\
\sigma_{t+1}^2 = \kappa + \xi \sigma_{t}^2(\eta_{t|t-1}-\lambda_t)^2 + \zeta\sigma_{t}^2
\end{cases},
\end{equation*}
where $\eta_{t+1|t} \sim \textup{stdNTS}(\alpha, \theta, B)$ with constant $B\in \R$ (See \cite{Kim_et_al:2010:JBF} and \cite{RachevKimBianchiFabozzi:2011a} for more details).
Daily risk free rate of return and daily dividend rate of S\&P 500 index of the day are
$r = 0.3841\textup{bp}$ and $d = 0.7148\textup{bp}$, respectively. GARCH parameters estimated historical S\&P 500 index return by 5/10/2017 are $(\zeta$, $\xi$, $\kappa)$ = $(0.7237 $, $0.1979 $, $4.9418\cdot 10^{-6})$. The volatility and residual of the day is $\sigma_0=0.0046$ and $\epsilon_0 = 0.1651$, respectively.
Calibrated standard NTS parameters of GARCH-NTS model are $(\alpha, \theta, B)$ $= (0.4936, 0.1077, -0.5926)$ and $\lambda = 0.5026$, while calibrated parameters of the StoT-NTS model are 
$(\alpha$, $\theta$, $a_B$, $\sigma_B$, $B_0)$ $=(0.4638$, $0.1109$, $0.2513$, $0.0317$, $-0.6584)$ and  $\lambda = 0.5304$. Other calibrated parameters for the second Wednesday of each month from January 2016 to December 2017 are presented in Table \ref{Table:CalibratedParameters}.

For the performance analysis, we use four error estimators,
the average absolute error (AAE), the average absolute error as a percentage of the mean price (APE), the average relative percentage error (ARPE), and the square root of mean square relative error (RMSRE), defined as follows,
\begin{align*}
    \textup{AAE} =
    \sum_{n=1}^N \frac{|{P}_n -
    \widehat{P}_n|}{N},
&
~~~ 
   \textup{APE} = \frac{\textup{AAE}}{\sum_{n=1}^N\frac{{P}_n}{N}},
    \\
    \textup{ARPE} = 
    \sum_{n=1}^N \frac{|{P}_n -
    \widehat{P}_n|}{N{P}_n},
&~~~ 
   \textup{RMSRE} = \sqrt{
    \sum_{n=1}^N \frac{({P}_n -
    \widehat{P}_n)^2}{N{P}_n^2}},
\end{align*}
where $\widehat{P}_n$ and ${P}_n$ are model prices and observed market prices of options (OTM calls or OTM puts) with strikes $K_n$, time to maturity $T_n$,  $n\in\{1,\ldots,N\}$, and $N$ is the number of observed prices.
Those four error estimators of the GARCH-NTS and the StoT-NTS models are presented in Table \ref{Table:ErrorEstimator}. According to the table, we can see that all the error estimators for the StoT-NTS model are less than corresponding error estimators of the GARCH-NTS model except the cases of 01/13/2016 and 4/12/2017, on which two model errors are similar. 

To verify that the StoT-NTS model performs better than the benchmark GARCH-NTS model, we perform the simple hypothesis tests for APE, and ARPE.
Since AAE and APE have the same t-statistic values, we do not need to test both, but we present only APE case test. RMSRE is also omitted in this hypothesis test, since it is not linear.
Instead of the hypothesis ``the StoT-NTS model performs better than the benchmark GARCH-NTS model'', we use the following equivalent hypothesis:
\[
H_0: \mu_{NTS} \le \mu_{StoT} ~~~\text{ vs }~~~ H_1: \mu_{NTS} > \mu_{StoT}
\]
or
\[
H_0: \mu_{NTS} - \mu_{StoT} \le 0~~~\text{ vs }~~~ H_1: \mu_{NTS} - \mu_{StoT}  > 0
\]
where $\mu_{StoT}$ and $\mu_{NTS}$ are means of calibration errors for the StoT-NTS model and  GARCH-NTS model, respectively.

Let $N$ be the number of observed prices.
Let ${P}_n$ be observed market prices of option and $\widehat{P}_n^{StoT}$ and $\widehat{P}_n^{NTS}$ be model prices for the StoT-NTS model and the GARCH-NTS model, respectively, for $n\in\{1,\ldots,N\}$. Then calibration errors are defined by
\begin{align*}
&e_{APE}^{StoT}(n) = \frac{|{P}_n - \widehat{P}_n^{StoT}|}{\sum_{n=1}^N\frac{{P}_n}{N}} ~~~\text{ and }~~~ e_{AAE}^{NTS}(n) = \frac{|{P}_n - \widehat{P}_n^{NTS}|}{\sum_{n=1}^N\frac{{P}_n}{N}} \\
&e_{ARPE}^{StoT}(n) = \frac{|{P}_n - \widehat{P}_n^{StoT}|}{{P}_n} ~~~\text{ and }~~~ e_{AAE}^{NTS}(n) = \frac{|{P}_n - \widehat{P}_n^{NTS}|}{{P}_n}  
\end{align*}
for APE, and ARPE, respectively.
We perform the t-test for the hypothesis test for the following two cases:
\begin{itemize}
\item For APE, we set $\mu_{StoT}=E[e_{APE}^{StoT}]$ and $\mu_{NTS}=E[e_{APE}^{NTS}]$.
Note that
$\mu_{StoT}$ and $\mu_{NTS}$ are APE's of the StoT-NTS model and the GARCH-NTS model, respectively.
\item For ARPE, we set $\mu_{StoT}=E[e_{ARPE}^{StoT}]$ and $\mu_{NTS}=E[e_{ARPE}^{NTS}]$.
Note that
$\mu_{StoT}$ and $\mu_{NTS}$ are ARPE's of the StoT-NTS model and the GARCH-NTS model, respectively.
\end{itemize}

The $t$-statistic and corresponding $p$-values of the tests in Table \ref{Table:PerformanceTest}. 
Except the date 01/12/2016, 01/11/2017, and 04/12/2017, $H_0$ is rejected for the APE case.
Except the date 01/12/2016, $H_0$ is rejected for the ARPE case. In the bottom line of the table, we present the result of those hypothesis tests for all market option prices and model prices we observed in this investigation. Considering total samples of every date in this investigation, $H_0$ is rejected for APE and ARPE.
We can conclude that the StoT-NTS model calibration performs typically better than the GARCH-NTS calibration in this investigation. 

\section{Conclusion}
In this paper, we present the StoT-NTS model obtained by taking a stochastic process for the parameter $B$ in NTS process. The model has the stochastic exponential tails, and it deduces the stochastic skewness and stochastic kurtosis of the residual of ARMA-GARCH-NTS model, and hence it captures the time-varying vol-of-vol of the stock or index return time series. Through the empirical test of S\&P 500 index return data, we observe that the skewness of the residual is typically negative. Also, if the skewness is decreasing, then the excess kurtosis of residual is increasing. The NTS distribution can describe this phenomenon by controlling the shape parameter $B$. By applying the ARIMA(1,1,0) model for the parameter $ B $ for time $t$, the StoT-NTS model describes the stochastic skewness and stochastic kurtosis empirically observed in S\&P 500 index return data. The StoT-NTS option pricing model is also discussed as an application of the StoT-NTS model. We present the Monte-Carlo simulation technique based on the model and calibrate the model to the S\&P 500 option prices observed in the market. In this empirical investigation, the StoT-NTS option pricing model performs mostly better than the benchmark GARCH-NTS option pricing model, since the former captures the time-varying vol-of-vol in the risk-neutral market, but the latter does not.

\singlespacing
\bibliographystyle{decsci_mod}
\bibliography{refs_aaron_StoT}
\clearpage

\begin{table}
\begin{footnotesize}
\begin{center}
\caption{\label{Table:GARCHparameters}GARCH Parameters, daily risk-free rate of return, and daily dividend rate}
\begin{tabular}{cccccccc}
Date & $\zeta$ & $\xi$ & $\kappa$ & $\sigma_0$ & $\epsilon_0$ & $r$ (bp) & $d$ (bp) \\
\hline
01/13/2016 & 
  $0.7328 $ & $0.1641 $ & $7.1125\cdot 10^{-6}$
  & $0.0108 $ & $-2.6226 $ & $0.1564 $ & $0.8696 $\\
 02/10/2016 & 
  $0.7396 $ & $0.1661 $ & $6.7564\cdot 10^{-6}$
  & $0.0119 $ & $-0.2969 $ & $0.1596 $ & $0.7830 $\\
 03/09/2016 & 
  $0.7502 $ & $0.1626 $ & $6.3634\cdot 10^{-6}$
  & $0.0088 $ & $0.5246 $ & $0.1601 $ & $0.9031 $\\
 04/13/2016 & 
  $0.7529 $ & $0.1601 $ & $6.2540\cdot 10^{-6}$
  & $0.0079 $ & $1.2029 $ & $0.1611 $ & $0.9105 $\\
 05/11/2016 & 
  $0.7410 $ & $0.1679 $ & $6.4898\cdot 10^{-6}$
  & $0.0076 $ & $-1.3254 $ & $0.1636 $ & $0.8990 $\\
 06/08/2016 & 
  $0.7308 $ & $0.1727 $ & $6.5207\cdot 10^{-6}$
  & $0.0056 $ & $0.4959 $ & $0.1653 $ & $0.9030 $\\
 07/06/2016 & 
  $0.6723 $ & $0.2044 $ & $8.5048\cdot 10^{-6}$
  & $0.0109 $ & $0.4533 $ & $0.1778 $ & $0.9033 $\\
 08/10/2016 & 
  $0.6882 $ & $0.1981 $ & $7.7233\cdot 10^{-6}$
  & $0.0058 $ & $-0.5718 $ & $0.1800 $ & $0.8830 $\\
 09/07/2016 & 
  $0.6723 $ & $0.2308 $ & $7.1132\cdot 10^{-6}$
  & $0.0052 $ & $-0.2410 $ & $0.1800 $ & $0.8500 $\\
 10/12/2016 & 
  $0.6537 $ & $0.2152 $ & $8.9111\cdot 10^{-6}$
  & $0.0083 $ & $0.0683 $ & $0.1852 $ & $0.8596 $\\
 11/09/2016 & 
  $0.6960 $ & $0.2035 $ & $6.8241\cdot 10^{-6}$
  & $0.0099 $ & $1.0756 $ & $0.1855 $ & $0.8332 $\\
 12/07/2016 & 
  $0.6957 $ & $0.2036 $ & $6.8541\cdot 10^{-6}$
  & $0.0056 $ & $2.2585 $ & $0.2002 $ & $0.7851 $\\
 01/11/2017 & 
  $0.6783 $ & $0.1998 $ & $8.0278\cdot 10^{-6}$
  & $0.0057 $ & $0.3904 $ & $0.2909 $ & $0.7621 $\\
 02/08/2017 & 
  $0.6961 $ & $0.1942 $ & $7.1116\cdot 10^{-6}$
  & $0.0055 $ & $0.0644 $ & $0.2919 $ & $0.8149 $\\
 03/08/2017 & 
  $0.6854 $ & $0.2051 $ & $7.1641\cdot 10^{-6}$
  & $0.0062 $ & $-0.4119 $ & $0.2981 $ & $0.7584 $\\
 04/12/2017 & 
  $0.7035 $ & $0.1936 $ & $6.4914\cdot 10^{-6}$
  & $0.0050 $ & $-0.9048 $ & $0.3841 $ & $0.6963 $\\
 05/10/2017 & 
  $0.7237 $ & $0.1979 $ & $4.9418\cdot 10^{-6}$
  & $0.0046 $ & $0.1651 $ & $0.3841 $ & $0.7148 $\\
 06/07/2017 & 
  $0.6954 $ & $0.1935 $ & $6.7354\cdot 10^{-6}$
  & $0.0055 $ & $0.1430 $ & $0.3952 $ & $0.7269 $\\
 07/05/2017 & 
  $0.7041 $ & $0.1899 $ & $6.3665\cdot 10^{-6}$
  & $0.0061 $ & $0.1588 $ & $0.4830 $ & $0.7274 $\\
 08/09/2017 & 
  $0.7042 $ & $0.2067 $ & $5.3908\cdot 10^{-6}$
  & $0.0046 $ & $-0.1730 $ & $0.4853 $ & $0.7466 $\\
 09/06/2017 & 
  $0.7075 $ & $0.2055 $ & $5.2737\cdot 10^{-6}$
  & $0.0062 $ & $0.4306 $ & $0.4845 $ & $0.7544 $\\
 10/11/2017 & 
  $0.6885 $ & $0.2128 $ & $5.8454\cdot 10^{-6}$
  & $0.0048 $ & $0.3022 $ & $0.4883 $ & $0.7023 $\\
 11/08/2017 & 
  $0.6990 $ & $0.2104 $ & $5.3214\cdot 10^{-6}$
  & $0.0045 $ & $0.2355 $ & $0.4885 $ & $0.6437 $\\
 12/06/2017 & 
  $0.6974 $ & $0.2113 $ & $5.3551\cdot 10^{-6}$
  & $0.0055 $ & $-0.1018 $ & $0.4980 $ & $0.6068 $\\
\hline 
\multicolumn{8}{r}{(bp $= 10^{-4}$)}
\end{tabular}
\end{center}
\end{footnotesize}
\end{table}

\begin{table}
\begin{footnotesize}
\begin{center}
\caption{\label{Table:CalibratedParameters}Calibrated Parameters}
\begin{tabular}{cccccccc}
 &  &\multicolumn{6}{c}{Parameters}\\
\cline{3-8}
Date  & Model & $\alpha$ & $\theta$ & $a_B$ & $\sigma_B$ & $B$ or $B_0$ & $\lambda$ \\
\hline
 01/13/2016 &  
GARCH-NTS & $0.3157 $ & $2.8308 $ & & & $-0.7533 $ & $0.6541 $\\
 & HuLK-NTS& $0.3157 $ & $2.8308 $ & $-0.9990 $ & $0.0039 $ & $-0.7543 $ & $0.6541 $\\
\hline 
 02/10/2016 &  
GARCH-NTS & $1.0877 $ & $17.3946 $ & & & $-0.5617 $ & $0.7547 $\\
 & HuLK-NTS& $1.0909 $ & $17.3939 $ & $0.0010 $ & $0.0533 $ & $-0.5808 $ & $0.7567 $\\
\hline 
 03/09/2016 &  
GARCH-NTS & $1.2678 $ & $4.5884 $ & & & $-0.9682 $ & $0.6730 $\\
 & HuLK-NTS& $1.2708 $ & $4.6373 $ & $0.2819 $ & $0.0096 $ & $-0.9898 $ & $0.6709 $\\
\hline 
 04/13/2016 &  
GARCH-NTS & $0.6467 $ & $0.4114 $ & & & $-0.8137 $ & $0.6016 $\\
 & HuLK-NTS& $0.7367 $ & $0.4107 $ & $0.8517 $ & $0.0020 $ & $-0.8305 $ & $0.6023 $\\
\hline 
 05/11/2016 &  
GARCH-NTS & $0.4776 $ & $0.8305 $ & & & $-0.8493 $ & $0.5565 $\\
 & HuLK-NTS& $0.4776 $ & $0.8305 $ & $0.2335 $ & $0.0065 $ & $-0.8675 $ & $0.5565 $\\
\hline 
 06/08/2016 &  
GARCH-NTS & $0.5757 $ & $0.5550 $ & & & $-0.9465 $ & $0.6194 $\\
 & HuLK-NTS& $0.5565 $ & $0.5703 $ & $0.8868 $ & $0.0028 $ & $-0.9507 $ & $0.6113 $\\
\hline 
 07/06/2016 &  
GARCH-NTS & $0.6141 $ & $0.4658 $ & & & $-0.8272 $ & $0.6163 $\\
 & HuLK-NTS& $0.6343 $ & $0.4300 $ & $-0.9963 $ & $0.0666 $ & $-0.8375 $ & $0.6067 $\\
\hline 
 08/10/2016 &  
GARCH-NTS & $0.0001 $ & $0.2165 $ & & & $-0.7442 $ & $0.5899 $\\
 & HuLK-NTS& $0.0001 $ & $0.1893 $ & $-0.0344 $ & $0.0322 $ & $-0.7478 $ & $0.6101 $\\
\hline 
 09/07/2016 &  
GARCH-NTS & $0.0004 $ & $0.4086 $ & & & $-0.8260 $ & $0.5021 $\\
 & HuLK-NTS& $0.0002 $ & $0.4305 $ & $0.1365 $ & $0.0152 $ & $-0.8523 $ & $0.4957 $\\
\hline 
 10/12/2016 &  
GARCH-NTS & $0.0002 $ & $0.9400 $ & & & $-0.9586 $ & $0.6031 $\\
 & HuLK-NTS& $0.0178 $ & $0.9569 $ & $-0.3504 $ & $0.0080 $ & $-0.9541 $ & $0.5976 $\\
\hline 
 11/09/2016 &  
GARCH-NTS & $0.0002 $ & $0.5519 $ & & & $-0.8848 $ & $0.5238 $\\
 & HuLK-NTS& $0.0002 $ & $0.5622 $ & $-0.0473 $ & $0.0138 $ & $-0.9000 $ & $0.5180 $\\
\hline 
 12/07/2016 &  
GARCH-NTS & $0.3802 $ & $0.1320 $ & & & $-0.4659 $ & $0.5585 $\\
 & HuLK-NTS& $0.5698 $ & $0.1237 $ & $-0.9961 $ & $0.0470 $ & $-0.5066 $ & $0.5708 $\\
\hline 
 01/11/2017 &  
GARCH-NTS & $0.0002 $ & $0.2608 $ & & & $-0.6486 $ & $0.4884 $\\
 & HuLK-NTS& $0.0002 $ & $0.2551 $ & $0.7315 $ & $0.0075 $ & $-0.6555 $ & $0.4945 $\\
\hline 
 02/08/2017 &  
GARCH-NTS & $0.3203 $ & $0.1201 $ & & & $-0.6169 $ & $0.5855 $\\
 & HuLK-NTS& $0.2157 $ & $0.1296 $ & $0.5971 $ & $0.0251 $ & $-0.7121 $ & $0.5985 $\\
\hline 
 03/08/2017 &  
GARCH-NTS & $0.9214 $ & $0.0687 $ & & & $-0.5017 $ & $0.5752 $\\
 & HuLK-NTS& $0.9214 $ & $0.0692 $ & $-0.9965 $ & $0.0268 $ & $-0.5066 $ & $0.5758 $\\
\hline 
 04/12/2017 &  
GARCH-NTS & $0.0001 $ & $0.5754 $ & & & $-0.9999 $ & $0.5935 $\\
 & HuLK-NTS& $0.0001 $ & $0.5936 $ & $-0.4362 $ & $0.0176 $ & $-0.9270 $ & $0.5828 $\\
\hline 
 05/10/2017 &  
GARCH-NTS & $0.4936 $ & $0.1077 $ & & & $-0.5926 $ & $0.5026 $\\
 & HuLK-NTS& $0.4638 $ & $0.1109 $ & $0.2513 $ & $0.0317 $ & $-0.6584 $ & $0.5304 $\\
\hline 
 06/07/2017 &  
GARCH-NTS & $0.4410 $ & $0.0518 $ & & & $-0.6267 $ & $0.6256 $\\
 & HuLK-NTS& $0.5993 $ & $0.0456 $ & $0.2350 $ & $0.0431 $ & $-0.7232 $ & $0.6690 $\\
\hline 
 07/05/2017 &  
GARCH-NTS & $0.4962 $ & $0.0571 $ & & & $-0.7152 $ & $0.6234 $\\
 & HuLK-NTS& $0.4753 $ & $0.0548 $ & $-0.3019 $ & $0.0395 $ & $-0.7358 $ & $0.6458 $\\
\hline 
 08/09/2017 &  
GARCH-NTS & $0.0002 $ & $0.1760 $ & & & $-0.8426 $ & $0.5959 $\\
 & HuLK-NTS& $0.0001 $ & $0.1698 $ & $-0.5737 $ & $0.0323 $ & $-0.8492 $ & $0.5968 $\\
\hline 
 09/06/2017 &  
GARCH-NTS & $0.0001 $ & $0.2577 $ & & & $-0.7896 $ & $0.5676 $\\
 & HuLK-NTS& $0.0001 $ & $0.2559 $ & $-0.0055 $ & $0.0280 $ & $-0.8015 $ & $0.5675 $\\
\hline 
 10/11/2017 &  
GARCH-NTS & $0.0001 $ & $0.0523 $ & & & $-0.7615 $ & $0.6511 $\\
 & HuLK-NTS& $0.0132 $ & $0.0502 $ & $0.1785 $ & $0.0408 $ & $-0.8713 $ & $0.6618 $\\
\hline 
 11/08/2017 &  
GARCH-NTS & $0.5200 $ & $0.0742 $ & & & $-0.6892 $ & $0.5915 $\\
 & HuLK-NTS& $0.5003 $ & $0.0619 $ & $-0.3800 $ & $0.0725 $ & $-0.7321 $ & $0.6341 $\\
\hline 
 12/06/2017 &  
GARCH-NTS & $1.1843 $ & $0.0293 $ & & & $-0.6654 $ & $0.7303 $\\
 & HuLK-NTS& $1.1843 $ & $0.0293 $ & $0.0969 $ & $0.0195 $ & $-0.6677 $ & $0.7303 $\\
\hline 
\end{tabular}
\end{center}
\end{footnotesize}
\end{table}

\begin{table}
\begin{footnotesize}
\begin{center}
\caption{\label{Table:ErrorEstimator}Error Estimators}
\begin{tabular}{cccccc}
Date & Model & AAE & APE & ARPE & RMSRE \\
\hline
 01/13/2016 & 
 NTS & $1.0433 $ & $0.0627 $ & $0.2086 $ & $0.3005 $\\
 & HuLK & $1.0439 $ & $0.0627 $ & $0.2090 $ & $0.3009 $\\
\hline 
 02/10/2016 & 
 NTS & $2.5294 $ & $0.1303 $ & $0.3708 $ & $0.7047 $\\
 & HuLK & $2.4862 $ & $0.1280 $ & $0.3532 $ & $0.6844 $\\
\hline 
 03/09/2016 & 
 NTS & $1.2637 $ & $0.0830 $ & $0.2717 $ & $0.4972 $\\
 & HuLK & $1.1922 $ & $0.0783 $ & $0.2535 $ & $0.4587 $\\
\hline 
 04/13/2016 & 
 NTS & $0.7991 $ & $0.0695 $ & $0.2946 $ & $0.4321 $\\
 & HuLK & $0.7871 $ & $0.0685 $ & $0.2919 $ & $0.4321 $\\
\hline 
 05/11/2016 & 
 NTS & $0.8853 $ & $0.0826 $ & $0.3687 $ & $0.6119 $\\
 & HuLK & $0.8786 $ & $0.0820 $ & $0.3747 $ & $0.6304 $\\
\hline 
 06/08/2016 & 
 NTS & $0.5778 $ & $0.0560 $ & $0.2260 $ & $0.3538 $\\
 & HuLK & $0.5225 $ & $0.0506 $ & $0.2019 $ & $0.3162 $\\
\hline 
 07/06/2016 & 
 NTS & $0.9900 $ & $0.0934 $ & $0.3577 $ & $0.5140 $\\
 & HuLK & $0.9160 $ & $0.0864 $ & $0.3331 $ & $0.4990 $\\
\hline 
 08/10/2016 & 
 NTS & $0.8431 $ & $0.0906 $ & $0.3438 $ & $0.5258 $\\
 & HuLK & $0.7956 $ & $0.0855 $ & $0.3164 $ & $0.4922 $\\
\hline 
 09/07/2016 & 
 NTS & $1.1065 $ & $0.1131 $ & $0.4689 $ & $0.7001 $\\
 & HuLK & $1.0700 $ & $0.1093 $ & $0.4357 $ & $0.6462 $\\
\hline 
 10/12/2016 & 
 NTS & $1.0079 $ & $0.0725 $ & $0.3171 $ & $0.6005 $\\
 & HuLK & $0.9926 $ & $0.0714 $ & $0.3096 $ & $0.5873 $\\
\hline 
 11/09/2016 & 
 NTS & $0.6946 $ & $0.0717 $ & $0.3003 $ & $0.4240 $\\
 & HuLK & $0.6644 $ & $0.0686 $ & $0.2909 $ & $0.4222 $\\
\hline 
 12/07/2016 & 
 NTS & $1.0412 $ & $0.0869 $ & $0.3368 $ & $0.5429 $\\
 & HuLK & $0.9970 $ & $0.0832 $ & $0.3301 $ & $0.5364 $\\
\hline 
 01/11/2017 & 
 NTS & $0.7766 $ & $0.0768 $ & $0.3218 $ & $0.4951 $\\
 & HuLK & $0.7651 $ & $0.0756 $ & $0.3025 $ & $0.4695 $\\
\hline 
 02/08/2017 & 
 NTS & $0.6380 $ & $0.0640 $ & $0.2337 $ & $0.4186 $\\
 & HuLK & $0.6057 $ & $0.0607 $ & $0.2134 $ & $0.3566 $\\
\hline 
 03/08/2017 & 
 NTS & $1.0550 $ & $0.0995 $ & $0.2997 $ & $0.4677 $\\
 & HuLK & $1.0349 $ & $0.0976 $ & $0.2794 $ & $0.4191 $\\
\hline 
 04/12/2017 & 
 NTS & $1.0937 $ & $0.0883 $ & $0.2558 $ & $0.4066 $\\
 & HuLK & $1.0966 $ & $0.0885 $ & $0.2219 $ & $0.3437 $\\
\hline 
 05/10/2017 & 
 NTS & $0.4560 $ & $0.0459 $ & $0.2194 $ & $0.4561 $\\
 & HuLK & $0.4002 $ & $0.0402 $ & $0.1880 $ & $0.3690 $\\
\hline 
 06/07/2017 & 
 NTS & $0.6847 $ & $0.0664 $ & $0.2270 $ & $0.3629 $\\
 & HuLK & $0.5811 $ & $0.0563 $ & $0.1876 $ & $0.2907 $\\
\hline 
 07/05/2017 & 
 NTS & $0.6090 $ & $0.0737 $ & $0.2887 $ & $0.4260 $\\
 & HuLK & $0.5764 $ & $0.0698 $ & $0.2540 $ & $0.3810 $\\
\hline 
 08/09/2017 & 
 NTS & $0.5904 $ & $0.0542 $ & $0.2203 $ & $0.4140 $\\
 & HuLK & $0.5601 $ & $0.0515 $ & $0.1980 $ & $0.3570 $\\
\hline 
 09/06/2017 & 
 NTS & $0.9301 $ & $0.0800 $ & $0.3284 $ & $0.5947 $\\
 & HuLK & $0.9006 $ & $0.0775 $ & $0.2821 $ & $0.4748 $\\
\hline 
 10/11/2017 & 
 NTS & $0.6265 $ & $0.0701 $ & $0.2331 $ & $0.3827 $\\
 & HuLK & $0.5492 $ & $0.0615 $ & $0.1759 $ & $0.2905 $\\
\hline 
 11/08/2017 & 
 NTS & $0.7517 $ & $0.0784 $ & $0.3576 $ & $0.6059 $\\
 & HuLK & $0.6257 $ & $0.0653 $ & $0.2810 $ & $0.4890 $\\
\hline 
 12/06/2017 & 
 NTS & $0.8312 $ & $0.0662 $ & $0.2492 $ & $0.5251 $\\
 & HuLK & $0.8156 $ & $0.0650 $ & $0.2286 $ & $0.4760 $\\
\hline 
\end{tabular}
\end{center}
\end{footnotesize}
\end{table}

\begin{sidewaystable}
\begin{footnotesize}
\begin{center}
\caption{\label{Table:PerformanceTest}Hypothesis test for APE and ARPE}
\begin{tabular}{c|c|ccc|ccc}
 & & \multicolumn{3}{c|}{APE}  & \multicolumn{3}{c}{ARPE}  
 \\
 \cline{3-5}  \cline{6-8} 
Date & $N$ & $\mu_{NTS}-\mu_{HuLK}$ & $t$-statistic & $p$-value & $\mu_{NTS}-\mu_{HuLK}$ & $t$-statistic & $p$-value \\
\hline
   01/13/2016 &  $372 $ 
 & $-0.0000 $ &  $-1.2180$ & $0.8884$ 
 & $-0.0004 $ &  $-2.3838$ & $0.9914$ 
 \\ 
 02/10/2016 &  $308 $ 
 & $0.0022^{***} $ &  $3.6422$ & $1.35 \cdot 10^{-4}$ 
 & $0.0176^{***} $ &  $7.4625$ & $4.24 \cdot 10^{-14}$ 
 \\ 
 03/09/2016 &  $297 $ 
 & $0.0047^{***} $ &  $9.2688$ & $0$ 
 & $0.0182^{***} $ &  $5.4127$ & $3.10 \cdot 10^{-8}$ 
 \\ 
 04/13/2016 &  $309 $ 
 & $0.0010^{***} $ &  $4.3800$ & $5.93 \cdot 10^{-6}$ 
 & $0.0027^{***} $ &  $3.5345$ & $2.04 \cdot 10^{-4}$ 
 \\ 
 05/11/2016 &  $345 $ 
 & $0.0006^{*} $ &  $1.7110$ & $0.0435$ 
 & $-0.0060 $ &  $-4.0520$ & $1.0000$ 
 \\ 
 06/08/2016 &  $316 $ 
 & $0.0054^{***} $ &  $7.3825$ & $7.76 \cdot 10^{-14}$ 
 & $0.0241^{***} $ &  $8.7411$ & $0$ 
 \\ 
 07/06/2016 &  $346 $ 
 & $0.0070^{***} $ &  $7.3198$ & $1.24 \cdot 10^{-13}$ 
 & $0.0246^{***} $ &  $4.3254$ & $7.61 \cdot 10^{-6}$ 
 \\ 
 08/10/2016 &  $292 $ 
 & $0.0051^{***} $ &  $5.5679$ & $1.29 \cdot 10^{-8}$ 
 & $0.0274^{***} $ &  $6.4771$ & $4.68 \cdot 10^{-11}$ 
 \\ 
 09/07/2016 &  $300 $ 
 & $0.0037^{***} $ &  $5.5828$ & $1.18 \cdot 10^{-8}$ 
 & $0.0331^{***} $ &  $8.8502$ & $0$ 
 \\ 
 10/12/2016 &  $288 $ 
 & $0.0011^{***} $ &  $3.6713$ & $1.21 \cdot 10^{-4}$ 
 & $0.0075^{***} $ &  $3.8506$ & $5.89 \cdot 10^{-5}$ 
 \\ 
 11/09/2016 &  $423 $ 
 & $0.0031^{***} $ &  $6.9092$ & $2.44 \cdot 10^{-12}$ 
 & $0.0094^{***} $ &  $3.7933$ & $7.43 \cdot 10^{-5}$ 
 \\ 
 12/07/2016 &  $351 $ 
 & $0.0037^{***} $ &  $6.4276$ & $6.48 \cdot 10^{-11}$ 
 & $0.0067^{*} $ &  $1.6583$ & $0.0486$ 
 \\ 
 01/11/2017 &  $301 $ 
 & $0.0011 $ &  $1.5756$ & $0.0576$ 
 & $0.0193^{***} $ &  $6.9910$ & $1.37 \cdot 10^{-12}$ 
 \\ 
 02/08/2017 &  $203 $ 
 & $0.0032^{*} $ &  $2.1904$ & $0.0142$ 
 & $0.0204^{**} $ &  $2.9323$ & $1.68 \cdot 10^{-3}$ 
 \\ 
 03/08/2017 &  $306 $ 
 & $0.0019^{***} $ &  $4.3900$ & $5.67 \cdot 10^{-6}$ 
 & $0.0203^{***} $ &  $4.5583$ & $2.58 \cdot 10^{-6}$ 
 \\ 
 04/12/2017 &  $282 $ 
 & $-0.0002 $ &  $-0.1981$ & $0.5785$ 
 & $0.0339^{***} $ &  $5.5053$ & $1.84 \cdot 10^{-8}$ 
 \\ 
 05/10/2017 &  $259 $ 
 & $0.0056^{***} $ &  $5.1967$ & $1.01 \cdot 10^{-7}$ 
 & $0.0314^{***} $ &  $4.3879$ & $5.72 \cdot 10^{-6}$ 
 \\ 
 06/07/2017 &  $225 $ 
 & $0.0101^{***} $ &  $8.1728$ & $1.11 \cdot 10^{-16}$ 
 & $0.0394^{***} $ &  $5.9808$ & $1.11 \cdot 10^{-9}$ 
 \\ 
 07/05/2017 &  $321 $ 
 & $0.0040^{***} $ &  $5.1178$ & $1.55 \cdot 10^{-7}$ 
 & $0.0348^{***} $ &  $6.8810$ & $2.97 \cdot 10^{-12}$ 
 \\ 
 08/09/2017 &  $279 $ 
 & $0.0028^{***} $ &  $4.9184$ & $4.36 \cdot 10^{-7}$ 
 & $0.0224^{***} $ &  $4.9254$ & $4.21 \cdot 10^{-7}$ 
 \\ 
 09/06/2017 &  $272 $ 
 & $0.0025^{**} $ &  $2.7741$ & $2.77 \cdot 10^{-3}$ 
 & $0.0462^{***} $ &  $5.2929$ & $6.02 \cdot 10^{-8}$ 
 \\ 
 10/11/2017 &  $269 $ 
 & $0.0087^{***} $ &  $6.5538$ & $2.81 \cdot 10^{-11}$ 
 & $0.0573^{***} $ &  $7.7381$ & $5.00 \cdot 10^{-15}$ 
 \\ 
 11/08/2017 &  $271 $ 
 & $0.0131^{***} $ &  $8.8339$ & $0$ 
 & $0.0765^{***} $ &  $8.0176$ & $5.55 \cdot 10^{-16}$ 
 \\ 
 12/06/2017 &  $260 $ 
 & $0.0012^{*} $ &  $1.9172$ & $0.0276$ 
 & $0.0207^{***} $ &  $4.5108$ & $3.23 \cdot 10^{-6}$ 
 \\ 
  \hline 
 total & $7195 $
 & $0.0034^{***} $ &  $22.1464$ & $0$ 
 & $0.0231^{***} $ &  $23.5084$ & $0$ \\
 \hline 
\end{tabular}
\end{center}
\end{footnotesize}
\end{sidewaystable}

\end{document}